\documentclass[%
 aip, apl,
 amsmath,amssymb,
reprint,%
]{revtex4-2}
\usepackage{xcolor}
\usepackage{graphicx}
\usepackage{dcolumn}
\usepackage{enumitem}
\usepackage{balance}
\usepackage{flushend}
\usepackage{mathptmx,mathtools}
\usepackage[cmbraces,varbb,varvw]{newtxmath}
\usepackage{etoolbox}
\usepackage[scaled=0.9]{helvet}
\usepackage{titlesec}
\usepackage[colorlinks = true, linkcolor = black, citecolor =black, urlcolor = black, linkbordercolor = white, bookmarks = false, pdfstartview={XYZ null null 1.00}]{hyperref}
\usepackage{fancyhdr}
\titleformat{\section}
  {\normalfont\sffamily\large\bfseries\color{black}}
  {\thesection.}{1em}{}
\titleformat{\subsection}
  {\normalfont\sffamily\large\bfseries\color{black}}
  {\Alph{subsection}.}{1em}{}

\makeatletter
\def\@email#1#2
{%
 \endgroup
 \patchcmd{\titleblock@produce}
  {\frontmatter@RRAPformat}
  {\frontmatter@RRAPformat{\produce@RRAP{*#1\href{mailto:#2}{#2}}}\frontmatter@RRAPformat}
  {}{}
}%

\makeatletter
\let\@msm@th@eqref\eqref
\renewcommand{\eqref}[1]{%
  \begingroup
 \leavevmode
  \color{blue}%
  \hypersetup{linkbordercolor=[named]{blue}}%
  \@msm@th@eqref{#1}%
  \endgroup
}

\usepackage{filecontents}

\makeatother
\begin{document}

\preprint{AIP/123-QED}
\thispagestyle{empty} 
\begin{figure}
   \begin{center}
	\includegraphics[scale=0.80]{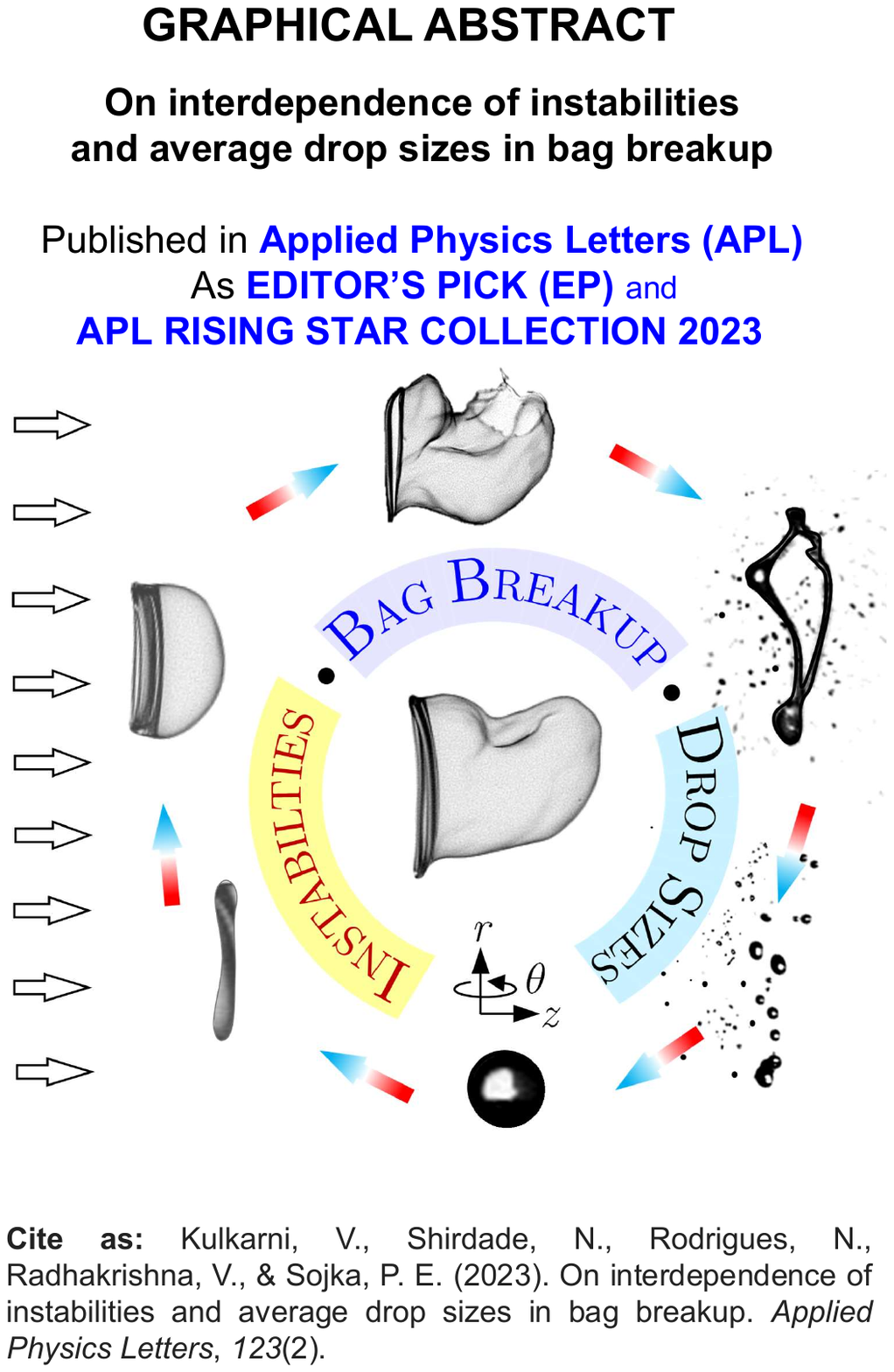}
   \end{center}
\end{figure}
\clearpage

\title[\textit{\footnotesize Applied Physics Letters}]{\textcolor{black}{On interdependence of instabilities and average drop sizes in bag breakup} \vspace{10pt}\\}
\author{Varun Kulkarni}
\altaffiliation[Corresponding author, electronic mail: ]{\href{mailto:varun14kul@gmail.com}{varun14kul@gmail.com}}
\affiliation{School of Mechanical Engineering, Purdue University, West Lafayette, IN 47906, \textcolor{black}{USA}}
\author{Nikhil Shirdade}
\affiliation{Department of Mechanical Engineering, Pennsylvania State University, State College, PA 16802 ,\textcolor{black}{USA} \looseness=-1} 
\author{Neil Rodrigues}
\affiliation{School of Mechanical Engineering, Purdue University, West Lafayette, IN 47906, \textcolor{black}{USA}}
\author{Vishnu Radhakrishna}
\affiliation{School of Mechanical Engineering, Purdue University, West Lafayette, IN 47906, \textcolor{black}{USA}}
\author{Paul E. Sojka}
\affiliation{School of Mechanical Engineering, Purdue University, West Lafayette, IN 47906, \textcolor{black}{USA}}



\begin{abstract}
A drop exposed to cross flow of air experiences sudden accelerations which deform it rapidly ultimately proceeding to disintegrate it into smaller fragments. In this work, we examine the breakup of a drop as a bag film with a bounding rim resulting from acceleration induced Rayleigh-Taylor instabilities and characterized through the Weber number, \textit{We}, representative of the competition between the disruptive aerodynamic force imparting acceleration and the restorative surface tension force. Our analysis reveals a previously overlooked parabolic dependence ($\sim We^2$) of the combination of dimensionless instability wavelengths $({\bar{\lambda}}_{bag}^2/ {\bar{\lambda}}_{rim}^4 {\bar{\lambda}}_{film})$ developing on different segments of the deforming drop. Further, we extend these findings to deduce the dependence of the average dimensionless drop sizes for the rim, $\langle {\widebar{D}}_{rim} \rangle$ and bag film, $\langle {\widebar{D}}_{film} \rangle$ individually, on $We$ and see them to decrease linearly for the rim ($\sim We^{-1}$) and quadratically for the bag film ($\sim We^{-2}$). The reported work is expected to have far-reaching implications as it provides unique insights on destabilization and disintegration mechanisms based on theoretical scaling arguments involving the commonly encountered canonical geometries of a toroidal rim and a curved liquid film.
\\ Area: Interdisciplinary Applied Physics, Surfaces and Interfaces.
\end{abstract}

\maketitle

Atomization of a single drop impacted by a stream of air appears as the constituent element of diverse natural and industrial processes, examples of which can be seen in falling raindrops\cite{Villermaux2009}, sea aerosols\cite{Troitskaya2018}, sneeze ejecta \cite{Scharfman2016} and, liquid propellant combustion.\cite{Wei2020} Of notable importance in this process is the deformation that the drop undergoes en route to its complete fragmentation. Depending on the air velocity, the morphology exhibited by the deforming drop could vary, leading to different regimes of breakup. \cite{Guildenbecher2009} One of these is the topological transition of a drop into a near-hemispherical \textit{bag} bounded by a toroidal \textit{rim} termed as \textit{bag breakup} and is significant as it sets the lower limit for air velocity at which complete drop fragmentation is assured.\cite{Kulkarni2014a, Kulkarni2014b} To characterize features of this regime across drops of different initial diameter, $D_0$ (in \textit{m}) and surface tension, $\sigma$ (in \textit{N}$\cdot$\textit{m}\textsuperscript{-1}) the balance between the restoring capillary pressure, $\sigma /D_0$ and opposing aerodynamic pressure, $\rho_a U_a^2$ is used to form a dimensionless number known as the Weber number,\cite{Sharma2022, Kulkarni2014a, Guildenbecher2009}\textcolor{black}{$We \left(= \rho_a U_a^2 D_0 /\sigma \right)$}, where, $\rho_a$ is the air density (in \textit{kg}$\cdot$\textit{m}\textsuperscript{-3}) and $U_a$ is the air velocity (in \textit{m}$\cdot$\textit{s}\textsuperscript{-1}). In these terms, bag breakup in the presence of a continuous cross stream of air is found to occur between $12 \lessapprox We \lessapprox 20$. \cite{Kulkarni2014a, Krzeczkowski1980, Jain2015, Guildenbecher2009, Chryssakis2008}

\begin{figure*}
   \vspace{0pt}
	\centerline{\includegraphics[width=\textwidth]{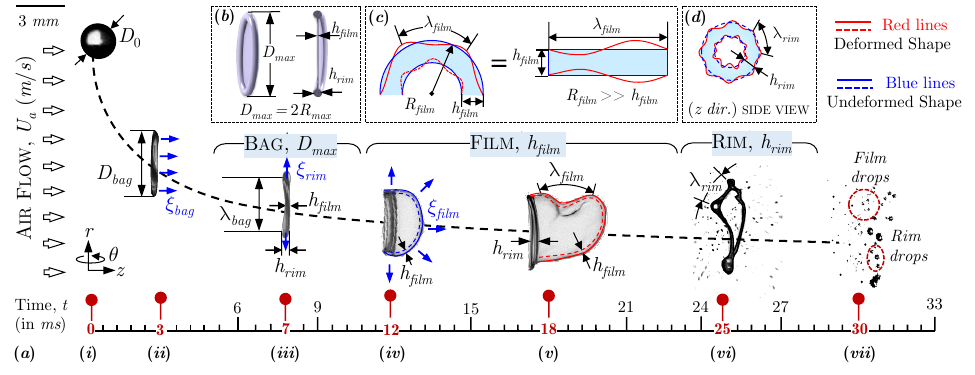}}
   \vspace{0pt}
   \caption{(a) (i) Initial undeformed drop (diameter, $D_0$) before it enters the air stream blowing from left to right and shown by hollow block arrows (ii) Flattening of the drop into a disc-like shape of cross stream dimension, $D_{bag}$ due to air impact and flattening in the radial (\textit{r}) and streamwise (\textit{z}) direction (iii) Incipience of Rayleigh-Taylor instability on $D_{bag}$ due streamwise acceleration, $\xi_{bag}$ and separation of drop into two distinct length scales, $h_{rim}$, and $h_{film}$ with $h_{rim}$ also accelerated radially, $\xi_{rim}$ (shown by blue arrows) (iv) Acceleration, $\xi_{film}$ of finite thickness ($h_{film}$) curved film (shown by blue arrows) (v) Instabilities on curved bag film due to radial acceleration, $\xi_{film}$ (vi) Bursting of curved liquid bag film (vii) Fragmentation of rim (b) 3D rendering of discoid (left) with the cut section (right) showing the different length scales, $h_{rim}$, \textcolor{black}{$D_{bag}$} and $h_{film}$ (c) Illustration showing the approximation of the unstable curved film to one with flat interfaces since radius of the film, $\textcolor{black}{R_{film}} >> h_{rim}$, refer (a)-(v) (d) Schematic showing waves on rim, $\lambda_{rim}$ from \textit{z} direction, refer (a)-(vi). Image sequence corresponds to $We \approx 16$ and 40\% by weight glycerine-water solution. Dashed black line shows the trajectory of the deforming drop.}
   \label{Fig1}
   \vspace{-15pt}
\end{figure*}

The morphological evolution of a drop as it undergoes bag breakup (see Fig. \ref{Fig1}) from (\textit{i}) its initial undeformed spherical shape, can be briefly described by its following five main stages\cite{Kulkarni2014a, Kulkarni2013,Wang2014} (\textit{ii}) Initial flattening of the drop (\textit{iii}) Formation of thin central film and rim (\textit{iv}) Growth of thin central film into a bag (\textit{v}) Bursting of curved film into drops (\textit{vi}) Disintegration of the rim into drops. Whilst most studies have focused on initial deformation \citep{Wang2014, Joshi2022, Opfer2014, Quan2006} (\textit{ii}) and the final disintegration\citep{Jackiw2022, Villermaux2009, Zhao2011b} (\textit{v, vi}), details of the connection between stages (\textit{iii} and \textit{iv}), governed by hydrodynamic instabilities and their influence on fragment production namely, stages (\textit{v} and \textit{vi}) has remained unexplored. These instabilities destabilize the liquid drop-air interface \cite{Zhao2011a, Zhao2011b, Zhao2010} in accordance with the Rayleigh-Taylor instability, known to occur when a heavy fluid (liquid drop), accelerates into a lighter fluid (air) \cite{Taylor1950, Rayleigh1882} thereby controlling the ensuing breakup dynamics. In this letter, we investigate the details of such manifestations and in so doing unearth a yet unreported, intriguing correlation between the three main instability wavelengths ($\lambda_{rim}$, $\lambda_{bag}$ and $\lambda_{film}$) which develop as the drop deforms and extend those findings to examine the differences in $We$ scaling for the average rim and film drops, $\langle D_{film}\rangle$ and $\langle D_{rim}\rangle$ respectively.

For our experiments, we use five test fluids which are four glycerine-water mixtures, 40\%, 50\%, 63\%, 70\% by weight and D.I. water. The fluids considered are incompressible with constant density, $\rho_l$ (in \textit{kg}$\cdot$\textit{m}\textsuperscript{-3}) and low dynamic viscosity, $\mu_l$ (in \textit{Pa}$\cdot$\textit{s}) which correspond to the dimensionless Ohnesorge number, $Oh = \mu_l/\sqrt{\rho_l \sigma D_0} < 0.1$ hence eliminating the role of viscosity in our findings and consistent with earlier studies.\cite{Jain2015,Soni2020,Zhao2011a} The apparatus and fluids used are similar to those used in our prior work \cite{Kulkarni2014a, Kulkarni2014b, Kulkarni2012a, Kulkarni2012b, Kulkarni2015} and consist of a converging nozzle producing a near flat air velocity profile. Drops ($D_0 \approx 2.1 - 2.6$ \textit{mm}) are released vertically from a syringe needle and are small enough to neglect gravitational effects. The air flow velocities are varied between 10 to 13 \textit{m}$\cdot$\textit{s}\textsuperscript{-1} which correspond to $12 \lessapprox We \lessapprox 20$ within the bag breakup regime. The experimental observations (see Fig.\ref{Fig1}\textcolor{black}{(a), (\textit{i})-(\textit{vii})}) are recorded using videos taken at 4700 fps and a resolution of 800 $\times$ 600 pixels such that 1 pixel $\approx$ 65 $\mu m$. It is worth mentioning that drop breakup and deformation in our work is kindred with studies using shock waves\cite{Theofanous2011, Hsiang1995, Sharma2022} although bag breakup is observed for a slightly broader range of $10 \lessapprox We \lessapprox 24 \textrm{ to } 35$ due to differences in aerodynamic loading. 

To measure bag and rim drop sizes, 3 videos for a given $We$ and $Oh$ were analyzed using thresholding and post processing described in literature.\cite{Zhao2011b, Villermaux2009} Film drops were measured once the bag bursting is complete while the rim is still intact (see Fig. \ref{Fig1}\textcolor{black}{(a), (\textit{v})}) and rim drops were measured upon complete disintegration of the rim (see Fig. \ref{Fig1}\textcolor{black}{(a), (\textit{v})}). Node formation in the rim was not considered dominant due to low  \textit{Oh} and comparatively lower $We$ compared to previous works. \cite{Zhao2010, Jackiw2022} The measured bag and rim drop sizes are an arithmetic average of all rim drops ($\approx$ 50) and majority of the bag film drops ($\approx$ 1000) at a given $We$ for the five test fluids and above the spatial resolution of our imaging and as demonstrated in previous studies.\cite{Gao2013, Zhao2011b, Radhakrishna2021, Jackiw2022}A pixel resolution of $\approx$ 65 $\mu m$ was found to be adequate since majority of the drop fragments fall above 65 $\mu m$ and we are able to finely resolve drop sizes above this value (also see section S1 of supplementary material for additional details on the image processing procedure and validation).The total volume from the film drops ($\forall_{film}$) and the rim drops ($\forall_{rim}$) at each condition was combined to validate that nearly all the original drop volume ($\forall_{drop}$) is recovered. Lastly, in view of our experimental data, we assume that both the rim and curved film drops are monodispersed.  

Fig. \ref{Fig1}\textcolor{black}{(a), (\textit{i})-(\textit{vii})} shows the evolution of the drop deformation as it moves downstream after it enters the air stream. It starts with the \textcolor{black}{flattening of the drop \citep{Yang2017, Jain2015} which makes it assume the form of a cylindrical disc of diameter,} \textcolor{black}{$D_{bag}$ (see sections S2 and S3 supplementary section for any details)} later transforming into a thin central circular sheet of thickness, $h_{film}$ bounded by a thicker rim, $h_{rim}$ as depicted in Fig. \ref{Fig1}\textcolor{black}{(b)}. \textcolor{black}{Experimentally, $D_{bag}$ is measured and identified at the stage when the drop is a near cylindrical disc (see Fig. \ref{Fig1}\textcolor{black}{(a) -(\textit{ii})}). It can also be demarcated approximately by the inner diameter of the bag once it forms initially. In the immediate moments after the disc structure deforms we see a distinct formation of a rim, at this stage $h_{rim}$ is identified as the thickness of the edge of the deformed drop (see Fig. \ref{Fig1}\textcolor{black}{(a) -(\textit{iii})}). Note that $D_{bag}$ is entirely different from $D_{max}$ (see Fig. \ref{Fig1}\textcolor{black}{(b)}) which is the maximum cross stream dimension and attained when the drop expands further radially. Lastly, $h_{film}$ is noted as the thickness of the central film when a distinct rim is formed and can also be approximated by the thickness of the bag when it bursts\citep{Opfer2014} (see Fig. \ref{Fig1}\textcolor{black}{(a) -(\textit{iv})}).} Such non-uniform flattening to form a rim and a central thin film is characteristic of drop impact phenomena where the surface tension acting along the circumference collects liquid in a bounding rim leaving behind a thin central region. \cite{Fang2022} Analogous to such impact phenomena is the expansion of holes on liquid sheets \cite{Savva2009, Lhuissier2012, Culick1960, Taylor1959} which displays similar features albeit, contrastingly, the expanding rim is surrounded by a thin liquid sheet rather than enclosing it. 

\textit{Interdependence of instabilities}: A direct consequence of the separation of length scales from $D_0$ to \textcolor{black}{$D_{bag}$}, $h_{rim}$ and $h_{film}$ is that each of these liquid segments is subjected to Rayleigh-Taylor instability individually, introduced by the same dynamic pressure, $\rho_a U_a^2$ of the oncoming air flow which drives the flattening of the drop with a velocity, $U_l$ developing an inertial stress in the drop of $\rho_l U_l^2$ such that\citep{Opfer2014}, $U_l = U_a \sqrt{\rho_a/\rho_l}$. This means that the accelerations experienced by each of the liquid segments can be deduced from their individual masses as determined by their respective length scales. We exploit this connection between the mass and acceleration in view of the fact that the wavelength of Rayleigh-Taylor instability strongly depends on the imposed acceleration (due to the dynamic air pressure) to establish a unique interdependence between the different wavelengths that develop on the drop, rim and, bag film as shown in Fig \ref{Fig1} \textcolor{black}{(a)-(d)}.

To begin we notice that as the drop expands radially reaching \textcolor{black}{the cross stream dimension, $D_{bag}$ (while still in its cylindrical disc form)} the acceleration experienced by it in the streamwise direction ($\xi_{bag}$) from an initial zero streamwise (\textit{z}) velocity to the air jet velocity, $U_a$ starts to manifest itself in the form of Rayleigh-Taylor waves, in their \textit{first} form (corresponding to blue arrows in Fig. \ref{Fig1}\textcolor{black}{(a), (\textit{ii})}). Characteristically, this is seen as a thick rim bounding a thin central film which later bulges into a bag signifying the amplitude of a wave of wavelength, $\lambda_{bag}$ (see Fig. \ref{Fig1}\textcolor{black}{(a), (\textit{iii}) and (\textit{iv})}). The rim of thickness, $h_{rim}$ formed in the process is radially accelerated (shown by blue arrows in Fig. \ref{Fig1}\textcolor{black}{(a), (\textit{iii})}) resulting in the emergence of waves of wavelength, $\lambda_{rim}$ (seen more prominently in Fig. \ref{Fig1}\textcolor{black}{(a), (\textit{vi})} and sketch in Fig. \ref{Fig1}\textcolor{black}{(d)}) on the periphery of the rim exhibiting the \textit{second} occurrence of Rayleigh-Taylor instability (also see section S2 of the supplementary material for more on experimental measurement details of $\lambda_{rim}$ and $\lambda_{bag}$).

By noting that surface tension enables mode selection in Rayleigh-Taylor instability,\cite{Pfeiffer2020, Zhao2011a,  Zhao2011b, Kulkarni2014a, Kulkarni2013} mathematically the most amplified wavelength in both the above scenarios scales as, $\sqrt{\sigma/\rho_l \xi}$ where $\xi$ (in \textit{m}$\cdot$\textit{s}\textsuperscript{-2}) is the acceleration imparted to the liquid surface. Individually, the accelerations in rim and bag are scale as, $\xi_{rim} \sim \textcolor{black}{U_l}^2/h_{rim}$ and $\xi_{bag} \sim \textcolor{black}{U_l}^2/\textcolor{black}{D_{bag}}$ respectively, using which we can write the following expressions for the most amplified wavelengths, made dimensionless using $D_0$ and represented by overbar $\overline{(\cdot)}$,
\begin{equation} \label{Eq1}
{\overline{\lambda}}_{rim} \sim \sqrt{\frac{h_{rim}}{D_0}}{We^{-1/2}}
\end{equation}
\begin{equation}\label{Eq2}
{\overline{\lambda}}_{bag} \sim \sqrt{\frac{D_{max}}{D_0}}{We^{-1/2}}
\end{equation}
\begin{figure}
   \vspace{0pt}
	\includegraphics[width=\columnwidth]{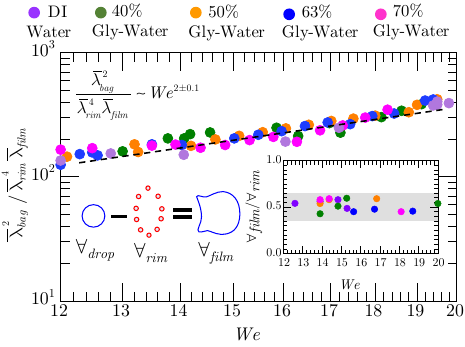}
   \vspace{-15pt}
   \caption{\textcolor{black}{Relation between the Rayleigh-Taylor instability dimensionless wavelengths on the drop (${\bar{\lambda}}_{bag}$), rim (${\bar{\lambda}}_{rim}$) and curved film (${\bar{\lambda}}_{film}$) with $We$ on log-log scale with an $R^2 \approx 0.97$. Inset shows $\forall_{rim}/ \forall_{film} \approx$ constant across all $We$ accompanied by a sketch to show accurate calculation of $\forall_{film}$ by considering difference between $\forall_{drop}$ and $\forall_{rim}$}.}
   \label{Fig2}
   \vspace{-10pt}
\end{figure}
As the bag expands the near-hemispherical curved thin liquid film is accelerated radially (see Fig. \ref{Fig1}\textcolor{black}{(a), (\textit{iv})}) leading to undulations of wavelength, $\lambda_{film}$ on its surface owing to Rayleigh-Taylor instability, now in its \textit{third} form and \textcolor{black}{experimentally measured just before the bag bursts}(see Fig. \ref{Fig1}\textcolor{black}{(a), (\textit{v})}). However, unlike the previous occurrences of the instability, here it acts on both the interfaces (inside and outside) of the film with finite thickness, $h_{film}$ approximated as a flat interface since $h_{film} <<  R_{film}$ (see sketch in Fig. \ref{Fig1}\textcolor{black}{(c)}) and $\lambda_{film} \gtrapprox h_{film}$. The existence of two interfaces stifles the development of the instability and the maximum dimensionless wavelength of instability\cite{Keller1954, Vledouts2016} for a film acceleration\citep{Villermaux2009} given by, $\xi_{film} \sim U_l^2/D_{0}$ now reads,
\begin{equation}\label{Eq3}
{\overline{\lambda}}_{film} \sim \left({\frac{h_{film}}{D_0}}\right)^{-1}{We^{-1}}
\end{equation} 
\begin{figure*}
   \vspace{0pt}
	\centerline{\includegraphics[width=\textwidth]{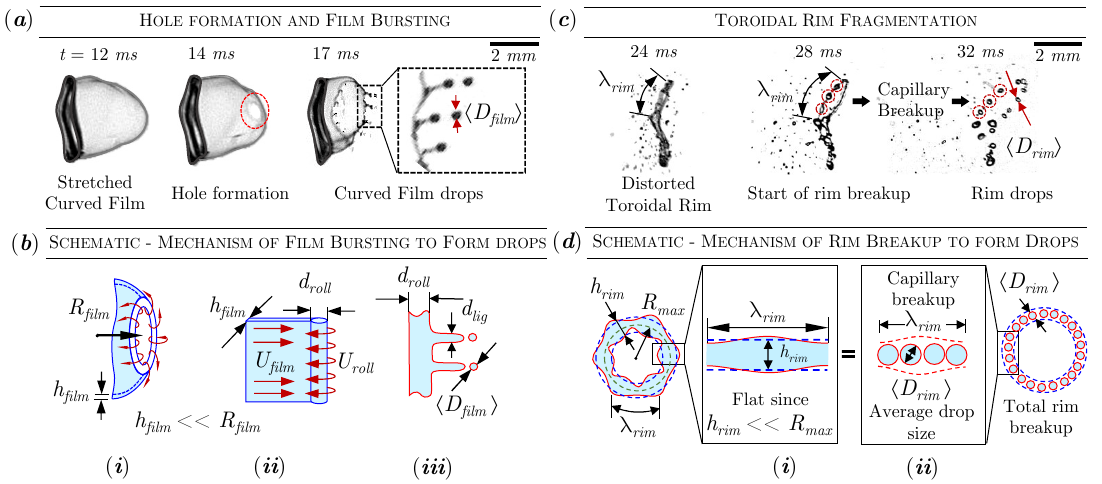}}
   \vspace{-10pt}
   \caption{(a) Initial bag (\textit{t} = 12 \textit{ms}) with formation of hole (\textit{t} = 14 \textit{ms}) triggering retraction of the curved film to form a ``rolling rim'' (\textit{t} = 17 \textit{ms}) which forms ligaments shedding droplets, zoomed view in dotted box (b) Schematic of curved bag film bursting leading to drop production (i) Initial curved film of radius, $R_{film}$ with a hole and rolling toroidal rim (ii) Approximation of the curved film as a flat sheet since $R_{film} >> h_{film}$ (iii) Formation of drops of size, $\langle D_{film}\rangle$ from ligaments of size, $d_{lig}$ produced from the rolling rim of diameter, $d_{roll}$ (c) Rim distortions of wavelength $\lambda_{rim}$ after bursting of the curved film (\textit{t} = 24 \textit{ms}) followed by emergence of bulbous regions due to capillary instability (\textit{t} = 28 \textit{ms}) finally resulting in a ring of spherical drops (\textit{t} = 32 \textit{ms}) (d) Schematic of rim breakup into drops (i) Approximation of toroidal ring as a flat interface since $h_{rim} <<  R_{max}$ (ii) Volume in one wavelength, $\lambda_{rim}$ equated to that of drops of average diameter, $\langle D_{rim}\rangle$. \textcolor{black}{Test conditions, 50\% glycerine-water, $We \approx 16$}.}
   \label{Fig3}
   \vspace{-15pt}
\end{figure*}

Since the drop has finite volume which is conserved at all times during its deformation, the distribution of volume in the rim, $\forall_{rim} (\sim h_{rim}^2 \textcolor{black}{D_{bag}}$) and the thin central disc-like sheet which later balloons into a bag, $\forall_{film} (\sim \textcolor{black}{D_{bag}}^2 h_{rim}$) equals the total drop volume, $\forall_{drop} (\sim d_0^3$). As confirmed by our experiments (see Fig. \ref{Fig2} inset), $\forall_{rim} \approx \alpha\forall_{film}$ \textcolor{black}{at all $We$} where, $\alpha \approx 0.5$ in our case and reported to be $\pm 0.2$ of this value elsewhere. Both $\forall_{rim}$ and $\forall_{film}$ \textcolor{black}{are independent\cite{Zhao2011b} of $We$ and recorded immediately when the deformed drop separates into a rim and a thin central film (see Fig. \ref{Fig1}\textcolor{black}{(a), (\textit{iii})})}. Heuristically, the radial expansion of the drop beyond the state its length is $D_{bag}$ also thins the toroidal rim, $h_{rim}$ and central region, $h_{film}$ such that volumes in the toroidal rim and thin central sheet are expected to be independently conserved \textcolor{black}{at any $We$. We provide further evidence for this in section S3 of supplementary material}. Such $We$ independent volume conservation is not uncommon and has been reported recently for drop impacts on solids.\citep{Wang2022} Therefore, we can write, $\forall_{rim} \sim \forall_{film}$ which for $\textcolor{black}{D_{bag}} >> h_{rim}$ can be expanded to, $h_{rim}^2 \textcolor{black}{D_{bag}} \sim \textcolor{black}{D_{bag}^2} h_{film}$ giving rise to the expression,
\begin{equation}\label{Eq3a}
h_{rim} \sim \sqrt{h_{film}\textcolor{black}{D_{bag}}}
\end{equation} 
Using eqn \ref{Eq3a}, we eliminate the different length scales $h_{rim}$, \textcolor{black}{$D_{bag}$} and $h_{film}$ in lieu of the expressions for different $\lambda$ in eqns \ref{Eq1}-\ref{Eq3} to obtain,
\begin{equation}\label{Eq4}
\frac{{{{\overline{\lambda}}_{bag}^2}}}{{\overline{\lambda}\;}_{rim}^4\;{\overline{\lambda}}_{film}} \sim We^2
\end{equation}
Fig. \ref{Fig2} shows the plot for eqn \ref{Eq4} overlayed with the experimental data with an \textcolor{black}{$R^2 \approx 0.97$}. \textcolor{black}{The scaling relation obtained here should be of immense significance to secondary atomization studies and can be extended to higher $Oh$ \citep{Radhakrishna2021} with appropriate modifications broadening the range of $We$ for which our theory is valid. Similarity, for low viscosity but large drops, $\mathcal{O}(cm)$ are results should be directly applicable. Unfortunately, no experimental data is available for such drops yet but existing numerical work\citep{Jalaal2012} has interestingly reported bag breakup at $We$ as large as 106 for such scenarios which we find to compare well with our findings (see supplementary section S2 for any additional details)}.

\textit{Mechanism of production of curved film drops}: The waves (of wavelength $\lambda_{film}$) mentioned so far do not grow indefinitely but reach their denouement once the curved film disintegrates into smaller drops. As the instabilities grow on this film it experiences thickness modulations along its two interfaces puncturing it at its rightmost tip where it is stretched the most (see Fig. \ref{Fig3}\textcolor{black}{(a), \textit{t} = 12 and 14 \textit{ms}}). The hole thus formed retracts collecting liquid mass in a ``rolling rim'' \citep{Culick1960, Taylor1959} as it moves along a curved path shown in Fig. \ref{Fig3}\textcolor{black}{(a), (\textit{iii})} and is accelerated centripetally which results in the emergence of waves \cite{Lhuissier2012} whose crests grow forming equally spaced ligaments culminating with the periodic shedding of droplets (see Fig. \ref{Fig3}\textcolor{black}{(a), \textit{t} = 17 \textit{ms}} and exploded view).

For predicting the average drop sizes (denoted by $\langle \cdot \rangle$) generated by this mechanism we assume that film drops of average diameter, $\langle{D}_{film}\rangle$ are produced from a ligament whose diameter, $d_{lig}$ equals that of the rolling rim, $d_{roll}$ which when expressed mathematically reads, $\langle{D}_{film}\rangle = d_{lig} = d_{roll}$. Hence, estimating $d_{roll}$ is sufficient to determine the film drop sizes. In order to do so, we assume that the mass accumulated in the rim (per unit width), $\frac{\pi}{4} d_{roll}^2$ in the time it takes to eject a single drop from the ligament, $\tau_{roll}$ is supplied by the flat liquid sheet (since $h_{film} << R_{film}$ ) which continually feeds this rim at a rate, $h_{film}U_{film}$ (see sketch in Fig. \ref{Fig3}\textcolor{black}{(b), (\textit{i})-(\textit{iii})}). Equating the two expressions gives us, $\langle{D}_{film}\rangle = d_{roll} = \sqrt{(4/\pi) h_{film} U_{film} \tau_{roll}}$. Here, $\tau_{roll} = h_{film}/U_{roll}$ and $U_{roll}$ is the velocity with the rim rolls and $U_{film}$ is the velocity with which the mass enters the rolling rim which are different since the liquid sheet stretches due to the air flow while the hole retracts unlike bursting bubbles.\cite{Lhuissier2012} Therefore, we need to evaluate only, $h_{film}$ and $U_{film}/U_{roll}$ to obtain $\langle{D}_{film}\rangle$ in terms of $We$. To this end, we balance the energy (per unit time \textcolor{black}{per unit area assuming unit width}) being provided by the moving sheet \textcolor{black}{of thickness $h_{film}$} written as, $(D_0^2 \rho_l U_{l}^2) U_{film}/h_{film}$ with that required to maintain the surface energy of the rim (\textcolor{black}{per unit time per unit area assuming unit width}), $\sigma U_{roll}$, resulting in the relation, $\frac{U_{film}}{U_{roll}} \sim \frac{\sigma h_{film}}{\rho_a U_a^2 D_0^2} = We^{-1}{\overline{h}}_{film}$, after noting that $\rho_l U_l^2 = \rho_a U_a^2$, as stated previously \textcolor{black}{(for more explanation on the details we refer the reader to section S4 of supplementary material)}. The other unknown ${\overline{h}}_{film}$ can be inferred from the observation that the expansion of the hole driven by the surface tension is resisted by its inertia  \cite{Lhuissier2012, Savva2009} giving rise to, $h_{film} \sim \sigma/\rho_l U_l^2$ which in dimensionless terms is, ${\overline{h}}_{film} \sim We^{-1}$ eventually allowing us to write, 
\begin{equation}\label{Eq5}
\langle\overline{D}_{film}\rangle \sim We^{-2}  
\end{equation}
\begin{figure}
   \vspace{-2pt}
	\centerline{\includegraphics[width=\columnwidth]{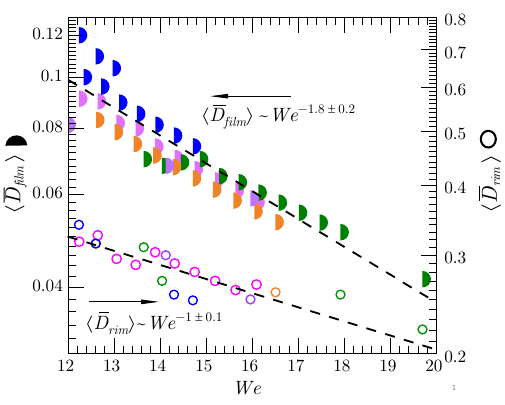}}
   \vspace{-10pt}
   \caption{Rim and bag drop sizes as a function $We$ derived theoretically in eqns \ref{Eq5} and \ref{Eq6} on semi-log scale. The average rim drops sizes, $\langle\overline{D}_{rim}\rangle$ decrease linearly whereas the average curved film sizes, $\langle\overline{D}_{film}\rangle$ decrease quadratically. Legend is same as Fig. \ref{Fig2}}
   \label{Fig4}
   \vspace{-15pt}
\end{figure}
\textit{Mechanism of production of rim drops}: Once the bag fragments completely as described above it leaves behind a destabilized free standing toroidal ring with distortions of wavelength, $\lambda_{rim}$ dictated by Rayleigh-Taylor instability which breaks up into drops owing to capillary (Rayleigh-Plateau) instability \citep{Villermaux2009, Jackiw2022} (see Fig. \ref{Fig3}\textcolor{black}{(c), \textit{t} = 24, 28} and \textcolor{black}{32 \textit{ms}}). The volume in each wavelength of the straight (since $h_{rim} << R_{max}$) segment, $\lambda_{rim} h_{rim}^2$ comprises of $N (\approx \textrm{4 or 5 at all \textit{We}})$ drops of average diameter, $\langle D_{rim} \rangle$ (see Fig. \ref{Fig3}\textcolor{black}{(d), (\textit{i}) and (\textit{ii})}). Further, we observe that inertia of the moving rim is opposed by the surface tension\citep{Opfer2014} such that, $h_{rim} \sim \sigma/\rho_l V_l^2$ which in its dimensionless form is, $\overline{h}_{rim} \sim We^{-1}$). Upon substituting for $h_{rim}$ and $\lambda_{rim}$ (from eqn \ref{Eq1}) in the scaling relation for volume conservation, $\langle D_{rim} \rangle \sim (\lambda_{rim} h_{rim}^2)^{1/3} $ we finally obtain,
\begin{equation}\label{Eq6}
\langle\overline{D}_{rim}\rangle \sim We^{-1}  
\end{equation}
The scaling relations derived in eqns \ref{Eq5}, \ref{Eq6} are plotted with the experimental data in Fig. \ref{Fig4} and a good match is seen. 

In conclusion, we have shown that during bag breakup, wavelength of instabilities on the flattened drop, thin film of the bag and the rim are related to each other. We also establish how the air flow affects droplet sizes in the aftermath of these instabilities. Implications of our results go beyond bag breakup, potentially providing crucial insights and points of comparison with several investigations on curved liquid films \cite{Mullagura2020, Naidu2022, Lhuissier2012, Vadivukkarasan2020} and toroidal rims \cite{Mehrabian2013, Pairam2009} which lead to droplet generation. Besides these, an interesting equivalence between studies on drop impact onto a deep liquid pool \citep{Kulkarni2021} to bag breakup has been propounded where the liquid and gas phases are inverted\citep{Opfer2014}, here too our findings could be of relevance. Given these similarities, our work should benefit industrial applications and fundamental studies alike.
\vspace{-20pt}
\section*{\normalsize Supplementary Material}
\vspace{-12pt}
The supplementary material accompanying this manuscript contains (\textit{i}) additional details leading to the scaling relations, (\ref{Eq4}), (\ref{Eq5}) and (\ref{Eq6}), (\textit{ii}) further justification for $We$ independent conservation of volumes of bag and rim and, (\textit{iii}) image processing steps followed to obtain rim and bag drop sizes and validation of those measurements.
\vspace{-20pt}
\section*{\normalsize Acknowledgment}
\vspace{-12pt}
Financial support for this project through the US Army Research office (ARO) under the Multi University Research Initiative (MURI) award number W911NF-08-l-0171 is gratefully acknowledged.
\vspace{-20pt}
\section*{\normalsize Data Availability Statement}
\vspace{-12pt}
Data underlying the conclusions of the paper are available in the plots presented and can be provided upon request.
\vspace{-20pt}
\section*{\normalsize Conflict of Interest Statement}
\vspace{-12pt}
The authors have no conflicts of interests to disclose.
\vspace{0pt}
\bibliography{Manuscript}

\begin{thebibliography}{47}%
\makeatletter
\providecommand \@ifxundefined [1]{%
 \@ifx{#1\undefined}
}%
\providecommand \@ifnum [1]{%
 \ifnum #1\expandafter \@firstoftwo
 \else \expandafter \@secondoftwo
 \fi
}%
\providecommand \@ifx [1]{%
 \ifx #1\expandafter \@firstoftwo
 \else \expandafter \@secondoftwo
 \fi
}%
\providecommand \natexlab [1]{#1}%
\providecommand \enquote  [1]{``#1''}%
\providecommand \bibnamefont  [1]{#1}%
\providecommand \bibfnamefont [1]{#1}%
\providecommand \citenamefont [1]{#1}%
\providecommand \href@noop [0]{\@secondoftwo}%
\providecommand \href [0]{\begingroup \@sanitize@url \@href}%
\providecommand \@href[1]{\@@startlink{#1}\@@href}%
\providecommand \@@href[1]{\endgroup#1\@@endlink}%
\providecommand \@sanitize@url [0]{\catcode `\\12\catcode `\$12\catcode
  `\&12\catcode `\#12\catcode `\^12\catcode `\_12\catcode `\%12\relax}%
\providecommand \@@startlink[1]{}%
\providecommand \@@endlink[0]{}%
\providecommand \url  [0]{\begingroup\@sanitize@url \@url }%
\providecommand \@url [1]{\endgroup\@href {#1}{\urlprefix }}%
\providecommand \urlprefix  [0]{URL }%
\providecommand \Eprint [0]{\href }%
\providecommand \doibase [0]{https://doi.org/}%
\providecommand \selectlanguage [0]{\@gobble}%
\providecommand \bibinfo  [0]{\@secondoftwo}%
\providecommand \bibfield  [0]{\@secondoftwo}%
\providecommand \translation [1]{[#1]}%
\providecommand \BibitemOpen [0]{}%
\providecommand \bibitemStop [0]{}%
\providecommand \bibitemNoStop [0]{.\EOS\space}%
\providecommand \EOS [0]{\spacefactor3000\relax}%
\providecommand \BibitemShut  [1]{\csname bibitem#1\endcsname}%
\let\auto@bib@innerbib\@empty
\bibitem [{\citenamefont {Villermaux}\ and\ \citenamefont
  {Bossa}(2009)}]{Villermaux2009}%
  \BibitemOpen
  \bibfield  {author} {\bibinfo {author} {\bibfnamefont {E.}~\bibnamefont
  {Villermaux}}\ and\ \bibinfo {author} {\bibfnamefont {B.}~\bibnamefont
  {Bossa}},\ }\bibfield  {title} {\enquote {\bibinfo {title} {Single-drop
  fragmentation determines size distribution of raindrops},}\ }\href@noop {}
  {\bibfield  {journal} {\bibinfo  {journal} {Nature Physics}\ }\textbf
  {\bibinfo {volume} {5}},\ \bibinfo {pages} {697--702} (\bibinfo {year}
  {2009})}\BibitemShut {NoStop}%
\bibitem [{\citenamefont {Troitskaya}\ \emph {et~al.}(2018)\citenamefont
  {Troitskaya}, \citenamefont {Druzhinin}, \citenamefont {Kozlov},\ and\
  \citenamefont {Zilitinkevich}}]{Troitskaya2018}%
  \BibitemOpen
  \bibfield  {author} {\bibinfo {author} {\bibfnamefont {Y.}~\bibnamefont
  {Troitskaya}}, \bibinfo {author} {\bibfnamefont {O.}~\bibnamefont
  {Druzhinin}}, \bibinfo {author} {\bibfnamefont {D.}~\bibnamefont {Kozlov}},\
  and\ \bibinfo {author} {\bibfnamefont {S.}~\bibnamefont {Zilitinkevich}},\
  }\bibfield  {title} {\enquote {\bibinfo {title} {The “bag breakup” spume
  droplet generation mechanism at high winds. part ii: Contribution to momentum
  and enthalpy transfer},}\ }\href@noop {} {\bibfield  {journal} {\bibinfo
  {journal} {Journal of Physical Oceanography}\ }\textbf {\bibinfo {volume}
  {48}},\ \bibinfo {pages} {2189--2207} (\bibinfo {year} {2018})}\BibitemShut
  {NoStop}%
\bibitem [{\citenamefont {Scharfman}\ \emph {et~al.}(2016)\citenamefont
  {Scharfman}, \citenamefont {Techet}, \citenamefont {Bush},\ and\
  \citenamefont {Bourouiba}}]{Scharfman2016}%
  \BibitemOpen
  \bibfield  {author} {\bibinfo {author} {\bibfnamefont {B.}~\bibnamefont
  {Scharfman}}, \bibinfo {author} {\bibfnamefont {A.}~\bibnamefont {Techet}},
  \bibinfo {author} {\bibfnamefont {J.}~\bibnamefont {Bush}},\ and\ \bibinfo
  {author} {\bibfnamefont {L.}~\bibnamefont {Bourouiba}},\ }\bibfield  {title}
  {\enquote {\bibinfo {title} {Visualization of sneeze ejecta: steps of fluid
  fragmentation leading to respiratory droplets},}\ }\href@noop {} {\bibfield
  {journal} {\bibinfo  {journal} {Experiments in Fluids}\ }\textbf {\bibinfo
  {volume} {57}},\ \bibinfo {pages} {1--9} (\bibinfo {year}
  {2016})}\BibitemShut {NoStop}%
\bibitem [{\citenamefont {Wei}\ \emph {et~al.}(2020)\citenamefont {Wei},
  \citenamefont {Yang}, \citenamefont {Zhang}, \citenamefont {Deng},
  \citenamefont {Liu}, \citenamefont {Law},\ and\ \citenamefont
  {Saha}}]{Wei2020}%
  \BibitemOpen
  \bibfield  {author} {\bibinfo {author} {\bibfnamefont {Y.}~\bibnamefont
  {Wei}}, \bibinfo {author} {\bibfnamefont {Y.}~\bibnamefont {Yang}}, \bibinfo
  {author} {\bibfnamefont {J.}~\bibnamefont {Zhang}}, \bibinfo {author}
  {\bibfnamefont {S.}~\bibnamefont {Deng}}, \bibinfo {author} {\bibfnamefont
  {S.}~\bibnamefont {Liu}}, \bibinfo {author} {\bibfnamefont {C.~K.}\
  \bibnamefont {Law}},\ and\ \bibinfo {author} {\bibfnamefont {A.}~\bibnamefont
  {Saha}},\ }\bibfield  {title} {\enquote {\bibinfo {title} {Atomization of
  acoustically levitated droplet exposed to hot gases},}\ }\href@noop {}
  {\bibfield  {journal} {\bibinfo  {journal} {Applied Physics Letters}\
  }\textbf {\bibinfo {volume} {116}},\ \bibinfo {pages} {044101} (\bibinfo
  {year} {2020})}\BibitemShut {NoStop}%
\bibitem [{\citenamefont {Guildenbecher}, \citenamefont {L{\'o}pez-Rivera},\
  and\ \citenamefont {Sojka}(2009)}]{Guildenbecher2009}%
  \BibitemOpen
  \bibfield  {author} {\bibinfo {author} {\bibfnamefont {D.}~\bibnamefont
  {Guildenbecher}}, \bibinfo {author} {\bibfnamefont {C.}~\bibnamefont
  {L{\'o}pez-Rivera}},\ and\ \bibinfo {author} {\bibfnamefont {P.}~\bibnamefont
  {Sojka}},\ }\bibfield  {title} {\enquote {\bibinfo {title} {Secondary
  atomization},}\ }\href@noop {} {\bibfield  {journal} {\bibinfo  {journal}
  {Experiments in Fluids}\ }\textbf {\bibinfo {volume} {46}},\ \bibinfo {pages}
  {371--402} (\bibinfo {year} {2009})}\BibitemShut {NoStop}%
\bibitem [{\citenamefont {Kulkarni}\ and\ \citenamefont
  {Sojka}(2014{\natexlab{a}})}]{Kulkarni2014a}%
  \BibitemOpen
  \bibfield  {author} {\bibinfo {author} {\bibfnamefont {V.}~\bibnamefont
  {Kulkarni}}\ and\ \bibinfo {author} {\bibfnamefont {P.~E.}\ \bibnamefont
  {Sojka}},\ }\bibfield  {title} {\enquote {\bibinfo {title} {Bag breakup of
  low viscosity drops in the presence of a continuous air jet},}\ }\href@noop
  {} {\bibfield  {journal} {\bibinfo  {journal} {Physics of Fluids}\ }\textbf
  {\bibinfo {volume} {26}},\ \bibinfo {pages} {072103} (\bibinfo {year}
  {2014}{\natexlab{a}})}\BibitemShut {NoStop}%
\bibitem [{\citenamefont {Kulkarni}\ and\ \citenamefont
  {Sojka}(2014{\natexlab{b}})}]{Kulkarni2014b}%
  \BibitemOpen
  \bibfield  {author} {\bibinfo {author} {\bibfnamefont {V.}~\bibnamefont
  {Kulkarni}}\ and\ \bibinfo {author} {\bibfnamefont {P.}~\bibnamefont
  {Sojka}},\ }\bibfield  {title} {\enquote {\bibinfo {title} {Fragmentation
  dynamics in the droplet bag breakup regime},}\ }in\ \href@noop {} {\emph
  {\bibinfo {booktitle} {APS Division of Fluid Dynamics Meeting Abstracts}}}\
  (\bibinfo {year} {2014})\ pp.\ \bibinfo {pages} {H12--001}\BibitemShut
  {NoStop}%
\bibitem [{\citenamefont {Sharma}\ \emph {et~al.}(2022)\citenamefont {Sharma},
  \citenamefont {Chandra}, \citenamefont {Basu},\ and\ \citenamefont
  {Kumar}}]{Sharma2022}%
  \BibitemOpen
  \bibfield  {author} {\bibinfo {author} {\bibfnamefont {S.}~\bibnamefont
  {Sharma}}, \bibinfo {author} {\bibfnamefont {N.~K.}\ \bibnamefont {Chandra}},
  \bibinfo {author} {\bibfnamefont {S.}~\bibnamefont {Basu}},\ and\ \bibinfo
  {author} {\bibfnamefont {A.}~\bibnamefont {Kumar}},\ }\bibfield  {title}
  {\enquote {\bibinfo {title} {Advances in droplet aerobreakup},}\ }\href@noop
  {} {\bibfield  {journal} {\bibinfo  {journal} {The European Physical Journal
  Special Topics}\ ,\ \bibinfo {pages} {1--15}} (\bibinfo {year}
  {2022})}\BibitemShut {NoStop}%
\bibitem [{\citenamefont {Krzeczkowski}(1980)}]{Krzeczkowski1980}%
  \BibitemOpen
  \bibfield  {author} {\bibinfo {author} {\bibfnamefont {S.~A.}\ \bibnamefont
  {Krzeczkowski}},\ }\bibfield  {title} {\enquote {\bibinfo {title}
  {Measurement of liquid droplet disintegration mechanisms},}\ }\href@noop {}
  {\bibfield  {journal} {\bibinfo  {journal} {International Journal of
  Multiphase Flow}\ }\textbf {\bibinfo {volume} {6}},\ \bibinfo {pages}
  {227--239} (\bibinfo {year} {1980})}\BibitemShut {NoStop}%
\bibitem [{\citenamefont {Jain}\ \emph {et~al.}(2015)\citenamefont {Jain},
  \citenamefont {Prakash}, \citenamefont {Tomar},\ and\ \citenamefont
  {Ravikrishna}}]{Jain2015}%
  \BibitemOpen
  \bibfield  {author} {\bibinfo {author} {\bibfnamefont {M.}~\bibnamefont
  {Jain}}, \bibinfo {author} {\bibfnamefont {R.~S.}\ \bibnamefont {Prakash}},
  \bibinfo {author} {\bibfnamefont {G.}~\bibnamefont {Tomar}},\ and\ \bibinfo
  {author} {\bibfnamefont {R.}~\bibnamefont {Ravikrishna}},\ }\bibfield
  {title} {\enquote {\bibinfo {title} {Secondary breakup of a drop at moderate
  weber numbers},}\ }\href@noop {} {\bibfield  {journal} {\bibinfo  {journal}
  {Proceedings of the Royal Society A: Mathematical, Physical and Engineering
  Sciences}\ }\textbf {\bibinfo {volume} {471}},\ \bibinfo {pages} {20140930}
  (\bibinfo {year} {2015})}\BibitemShut {NoStop}%
\bibitem [{\citenamefont {Chryssakis}\ and\ \citenamefont
  {Assanis}(2008)}]{Chryssakis2008}%
  \BibitemOpen
  \bibfield  {author} {\bibinfo {author} {\bibfnamefont {C.}~\bibnamefont
  {Chryssakis}}\ and\ \bibinfo {author} {\bibfnamefont {D.~N.}\ \bibnamefont
  {Assanis}},\ }\bibfield  {title} {\enquote {\bibinfo {title} {A unified fuel
  spray breakup model for internal combustion engine applications},}\
  }\href@noop {} {\bibfield  {journal} {\bibinfo  {journal} {Atomization and
  Sprays}\ }\textbf {\bibinfo {volume} {18}} (\bibinfo {year}
  {2008})}\BibitemShut {NoStop}%
\bibitem [{\citenamefont {Kulkarni}(2013)}]{Kulkarni2013}%
  \BibitemOpen
  \bibfield  {author} {\bibinfo {author} {\bibfnamefont {V.}~\bibnamefont
  {Kulkarni}},\ }\emph {\bibinfo {title} {An analytical and experimental study
  of secondary atomization of vibrational and bag breakup modes}},\ \href@noop
  {} {Ph.D. thesis},\ \bibinfo  {school} {Purdue University} (\bibinfo {year}
  {2013})\BibitemShut {NoStop}%
\bibitem [{\citenamefont {Wang}\ \emph {et~al.}(2014)\citenamefont {Wang},
  \citenamefont {Chang}, \citenamefont {Wu},\ and\ \citenamefont
  {Xu}}]{Wang2014}%
  \BibitemOpen
  \bibfield  {author} {\bibinfo {author} {\bibfnamefont {C.}~\bibnamefont
  {Wang}}, \bibinfo {author} {\bibfnamefont {S.}~\bibnamefont {Chang}},
  \bibinfo {author} {\bibfnamefont {H.}~\bibnamefont {Wu}},\ and\ \bibinfo
  {author} {\bibfnamefont {J.}~\bibnamefont {Xu}},\ }\bibfield  {title}
  {\enquote {\bibinfo {title} {Modeling of drop breakup in the bag breakup
  regime},}\ }\href@noop {} {\bibfield  {journal} {\bibinfo  {journal} {Applied
  Physics Letters}\ }\textbf {\bibinfo {volume} {104}},\ \bibinfo {pages}
  {154107} (\bibinfo {year} {2014})}\BibitemShut {NoStop}%
\bibitem [{\citenamefont {Joshi}\ and\ \citenamefont
  {Anand}(2022)}]{Joshi2022}%
  \BibitemOpen
  \bibfield  {author} {\bibinfo {author} {\bibfnamefont {S.}~\bibnamefont
  {Joshi}}\ and\ \bibinfo {author} {\bibfnamefont {T.}~\bibnamefont {Anand}},\
  }\bibfield  {title} {\enquote {\bibinfo {title} {Droplet deformation in
  secondary breakup: Transformation from a sphere to a disk-like structure},}\
  }\href@noop {} {\bibfield  {journal} {\bibinfo  {journal} {International
  Journal of Multiphase Flow}\ }\textbf {\bibinfo {volume} {146}},\ \bibinfo
  {pages} {103850} (\bibinfo {year} {2022})}\BibitemShut {NoStop}%
\bibitem [{\citenamefont {Opfer}\ \emph {et~al.}(2014)\citenamefont {Opfer},
  \citenamefont {Roisman}, \citenamefont {Venzmer}, \citenamefont
  {Klostermann},\ and\ \citenamefont {Tropea}}]{Opfer2014}%
  \BibitemOpen
  \bibfield  {author} {\bibinfo {author} {\bibfnamefont {L.}~\bibnamefont
  {Opfer}}, \bibinfo {author} {\bibfnamefont {I.}~\bibnamefont {Roisman}},
  \bibinfo {author} {\bibfnamefont {J.}~\bibnamefont {Venzmer}}, \bibinfo
  {author} {\bibfnamefont {M.}~\bibnamefont {Klostermann}},\ and\ \bibinfo
  {author} {\bibfnamefont {C.}~\bibnamefont {Tropea}},\ }\bibfield  {title}
  {\enquote {\bibinfo {title} {Droplet-air collision dynamics: Evolution of the
  film thickness},}\ }\href@noop {} {\bibfield  {journal} {\bibinfo  {journal}
  {Physical Review E}\ }\textbf {\bibinfo {volume} {89}},\ \bibinfo {pages}
  {013023} (\bibinfo {year} {2014})}\BibitemShut {NoStop}%
\bibitem [{\citenamefont {Quan}\ and\ \citenamefont
  {Schmidt}(2006)}]{Quan2006}%
  \BibitemOpen
  \bibfield  {author} {\bibinfo {author} {\bibfnamefont {S.}~\bibnamefont
  {Quan}}\ and\ \bibinfo {author} {\bibfnamefont {D.~P.}\ \bibnamefont
  {Schmidt}},\ }\bibfield  {title} {\enquote {\bibinfo {title} {Direct
  numerical study of a liquid droplet impulsively accelerated by gaseous
  flow},}\ }\href@noop {} {\bibfield  {journal} {\bibinfo  {journal} {Physics
  of Fluids}\ }\textbf {\bibinfo {volume} {18}},\ \bibinfo {pages} {102103}
  (\bibinfo {year} {2006})}\BibitemShut {NoStop}%
\bibitem [{\citenamefont {Jackiw}\ and\ \citenamefont
  {Ashgriz}(2022)}]{Jackiw2022}%
  \BibitemOpen
  \bibfield  {author} {\bibinfo {author} {\bibfnamefont {I.~M.}\ \bibnamefont
  {Jackiw}}\ and\ \bibinfo {author} {\bibfnamefont {N.}~\bibnamefont
  {Ashgriz}},\ }\bibfield  {title} {\enquote {\bibinfo {title} {Prediction of
  the droplet size distribution in aerodynamic droplet breakup},}\ }\href@noop
  {} {\bibfield  {journal} {\bibinfo  {journal} {Journal of Fluid Mechanics}\
  }\textbf {\bibinfo {volume} {940}} (\bibinfo {year} {2022})}\BibitemShut
  {NoStop}%
\bibitem [{\citenamefont {Zhao}\ \emph
  {et~al.}(2011{\natexlab{a}})\citenamefont {Zhao}, \citenamefont {Liu},
  \citenamefont {Xu},\ and\ \citenamefont {Li}}]{Zhao2011b}%
  \BibitemOpen
  \bibfield  {author} {\bibinfo {author} {\bibfnamefont {H.}~\bibnamefont
  {Zhao}}, \bibinfo {author} {\bibfnamefont {H.~F.}\ \bibnamefont {Liu}},
  \bibinfo {author} {\bibfnamefont {J.~L.}\ \bibnamefont {Xu}},\ and\ \bibinfo
  {author} {\bibfnamefont {W.~F.}\ \bibnamefont {Li}},\ }\bibfield  {title}
  {\enquote {\bibinfo {title} {Experimental study of drop size distribution in
  the bag breakup regime},}\ }\href@noop {} {\bibfield  {journal} {\bibinfo
  {journal} {Industrial \& Engineering Chemistry Research}\ }\textbf {\bibinfo
  {volume} {50}},\ \bibinfo {pages} {9767--9773} (\bibinfo {year}
  {2011}{\natexlab{a}})}\BibitemShut {NoStop}%
\bibitem [{\citenamefont {Zhao}\ \emph
  {et~al.}(2011{\natexlab{b}})\citenamefont {Zhao}, \citenamefont {Liu},
  \citenamefont {Cao}, \citenamefont {Li},\ and\ \citenamefont
  {Xu}}]{Zhao2011a}%
  \BibitemOpen
  \bibfield  {author} {\bibinfo {author} {\bibfnamefont {H.}~\bibnamefont
  {Zhao}}, \bibinfo {author} {\bibfnamefont {H.-F.}\ \bibnamefont {Liu}},
  \bibinfo {author} {\bibfnamefont {X.-K.}\ \bibnamefont {Cao}}, \bibinfo
  {author} {\bibfnamefont {W.-F.}\ \bibnamefont {Li}},\ and\ \bibinfo {author}
  {\bibfnamefont {J.-L.}\ \bibnamefont {Xu}},\ }\bibfield  {title} {\enquote
  {\bibinfo {title} {Breakup characteristics of liquid drops in bag regime by a
  continuous and uniform air jet flow},}\ }\href@noop {} {\bibfield  {journal}
  {\bibinfo  {journal} {International Journal of Multiphase Flow}\ }\textbf
  {\bibinfo {volume} {37}},\ \bibinfo {pages} {530--534} (\bibinfo {year}
  {2011}{\natexlab{b}})}\BibitemShut {NoStop}%
\bibitem [{\citenamefont {Zhao}\ \emph {et~al.}(2010)\citenamefont {Zhao},
  \citenamefont {Liu}, \citenamefont {Li},\ and\ \citenamefont
  {Xu}}]{Zhao2010}%
  \BibitemOpen
  \bibfield  {author} {\bibinfo {author} {\bibfnamefont {H.}~\bibnamefont
  {Zhao}}, \bibinfo {author} {\bibfnamefont {H.-F.}\ \bibnamefont {Liu}},
  \bibinfo {author} {\bibfnamefont {W.-F.}\ \bibnamefont {Li}},\ and\ \bibinfo
  {author} {\bibfnamefont {J.-L.}\ \bibnamefont {Xu}},\ }\bibfield  {title}
  {\enquote {\bibinfo {title} {Morphological classification of low viscosity
  drop bag breakup in a continuous air jet stream},}\ }\href@noop {} {\bibfield
   {journal} {\bibinfo  {journal} {Physics of Fluids}\ }\textbf {\bibinfo
  {volume} {22}},\ \bibinfo {pages} {114103} (\bibinfo {year}
  {2010})}\BibitemShut {NoStop}%
\bibitem [{\citenamefont {Taylor}(1950)}]{Taylor1950}%
  \BibitemOpen
  \bibfield  {author} {\bibinfo {author} {\bibfnamefont {G.~I.}\ \bibnamefont
  {Taylor}},\ }\bibfield  {title} {\enquote {\bibinfo {title} {The instability
  of liquid surfaces when accelerated in a direction perpendicular to their
  planes. i},}\ }\href@noop {} {\bibfield  {journal} {\bibinfo  {journal}
  {Proceedings of the Royal Society of London. Series A. Mathematical and
  Physical Sciences}\ }\textbf {\bibinfo {volume} {201}},\ \bibinfo {pages}
  {192--196} (\bibinfo {year} {\vspace{-1pt}1950})}\BibitemShut {NoStop}%
\bibitem [{\citenamefont {Rayleigh}(1882)}]{Rayleigh1882}%
  \BibitemOpen
  \bibfield  {author} {\bibinfo {author} {\bibfnamefont {R.}~\bibnamefont
  {Rayleigh}},\ }\bibfield  {title} {\enquote {\bibinfo {title} {Investigation
  of the character of the equilibrium of an incompressible heavy fluid of
  variable density},}\ }\href@noop {} {\bibfield  {journal} {\bibinfo
  {journal} {Proceedings of the London mathematical society}\ }\textbf
  {\bibinfo {volume} {1}},\ \bibinfo {pages} {170--177} (\bibinfo {year}
  {1882})}\BibitemShut {NoStop}%
\bibitem [{\citenamefont {Soni}\ \emph {et~al.}(2020)\citenamefont {Soni},
  \citenamefont {Kirar}, \citenamefont {Kolhe},\ and\ \citenamefont
  {Sahu}}]{Soni2020}%
  \BibitemOpen
  \bibfield  {author} {\bibinfo {author} {\bibfnamefont {S.~K.}\ \bibnamefont
  {Soni}}, \bibinfo {author} {\bibfnamefont {P.~K.}\ \bibnamefont {Kirar}},
  \bibinfo {author} {\bibfnamefont {P.}~\bibnamefont {Kolhe}},\ and\ \bibinfo
  {author} {\bibfnamefont {K.~C.}\ \bibnamefont {Sahu}},\ }\bibfield  {title}
  {\enquote {\bibinfo {title} {Deformation and breakup of droplets in an
  oblique continuous air stream},}\ }\href@noop {} {\bibfield  {journal}
  {\bibinfo  {journal} {International Journal of Multiphase Flow}\ }\textbf
  {\bibinfo {volume} {122}},\ \bibinfo {pages} {103141} (\bibinfo {year}
  {2020})}\BibitemShut {NoStop}%
\bibitem [{\citenamefont {Kulkarni}, \citenamefont {Guildenbecher},\ and\
  \citenamefont {Sojka}(2012)}]{Kulkarni2012a}%
  \BibitemOpen
  \bibfield  {author} {\bibinfo {author} {\bibfnamefont {V.}~\bibnamefont
  {Kulkarni}}, \bibinfo {author} {\bibfnamefont {D.}~\bibnamefont
  {Guildenbecher}},\ and\ \bibinfo {author} {\bibfnamefont {P.}~\bibnamefont
  {Sojka}},\ }\bibfield  {title} {\enquote {\bibinfo {title} {Secondary
  atomization of newtonian liquids in the bag breakup regime: Comparison of
  model predictions to experimental data},}\ }in\ \href@noop {} {\emph
  {\bibinfo {booktitle} {ICLASS 2012, 12th International Conference on Liquid
  Atomization and Spray Systems, Heidelberg, Germany}}}\ (\bibinfo {year}
  {2012})\BibitemShut {NoStop}%
\bibitem [{\citenamefont {Kulkarni}\ \emph {et~al.}(2012)\citenamefont
  {Kulkarni}, \citenamefont {Guildenbecher}, \citenamefont {Firehammer},\ and\
  \citenamefont {Sojka}}]{Kulkarni2012b}%
  \BibitemOpen
  \bibfield  {author} {\bibinfo {author} {\bibfnamefont {V.}~\bibnamefont
  {Kulkarni}}, \bibinfo {author} {\bibfnamefont {D.}~\bibnamefont
  {Guildenbecher}}, \bibinfo {author} {\bibfnamefont {S.}~\bibnamefont
  {Firehammer}},\ and\ \bibinfo {author} {\bibfnamefont {P.}~\bibnamefont
  {Sojka}},\ }\bibfield  {title} {\enquote {\bibinfo {title} {Bag breakup \vspace{-1pt}of
  viscous drops},}\ }in\ \href@noop {} {\emph {\bibinfo {booktitle} {APS
  Division of Fluid Dynamics Meeting Abstracts}}}\ (\bibinfo {year} {2012})\
  pp.\ \bibinfo {pages} {L8--001}\BibitemShut {NoStop}%
\bibitem [{\citenamefont {Kulkarni}\ \emph {et~al.}(2015)\citenamefont
  {Kulkarni}, \citenamefont {Bulusu}, \citenamefont {Plesniak},\ and\
  \citenamefont {Sojka}}]{Kulkarni2015}%
  \BibitemOpen
  \bibfield  {author} {\bibinfo {author} {\bibfnamefont {V.}~\bibnamefont
  {Kulkarni}}, \bibinfo {author} {\bibfnamefont {K.~V.}\ \bibnamefont
  {Bulusu}}, \bibinfo {author} {\bibfnamefont {M.~W.}\ \bibnamefont
  {Plesniak}},\ and\ \bibinfo {author} {\bibfnamefont {P.~E.}\ \bibnamefont
  {Sojka}},\ }\bibfield  {title} {\enquote {\bibinfo {title} {\vspace{-1pt}Fragment size
  distribution in viscous bag breakup of a drop},}\ }in\ \href@noop {} {\emph
  {\bibinfo {booktitle} {APS Division of Fluid Dynamics Meeting Abstracts}}}\
  (\bibinfo {year} {2015})\ pp.\ \bibinfo {pages} {D32--007}\BibitemShut
  {NoStop}%
\bibitem [{\citenamefont {Theofanous}(2011)}]{Theofanous2011}%
  \BibitemOpen
  \bibfield  {author} {\bibinfo {author} {\bibfnamefont {T.}~\bibnamefont
  {Theofanous}},\ }\bibfield  {title} {\enquote {\bibinfo {title} {Aerobreakup
  of newtonian and viscoelastic liquids},}\ }\href@noop {} {\bibfield
  {journal} {\bibinfo  {journal} {Annual Review of Fluid Mechanics}\ }\textbf
  {\bibinfo {volume} {43}},\ \bibinfo {pages} {661--690} (\bibinfo {year}
  {2011})}\BibitemShut {NoStop}%
\bibitem [{\citenamefont {Hsiang}\ and\ \citenamefont
  {Faeth}(1995)}]{Hsiang1995}%
  \BibitemOpen
  \bibfield  {author} {\bibinfo {author} {\bibfnamefont {L.-P.}\ \bibnamefont
  {Hsiang}}\ and\ \bibinfo {author} {\bibfnamefont {G.~M.}\ \bibnamefont
  {Faeth}},\ }\bibfield  {title} {\enquote {\bibinfo {title} {Drop deformation
  and breakup due to shock wave and steady disturbances},}\ }\href@noop {}
  {\bibfield  {journal} {\bibinfo  {journal} {International Journal of
  Multiphase Flow}\ }\textbf {\bibinfo {volume} {21}},\ \bibinfo {pages}
  {545--560} (\bibinfo {year} {1995})}\BibitemShut {NoStop}%
\bibitem [{\citenamefont {Gao}\ \emph {et~al.}(2013)\citenamefont {Gao},
  \citenamefont {Guildenbecher}, \citenamefont {Reu}, \citenamefont {Kulkarni},
  \citenamefont {Sojka},\ and\ \citenamefont {Chen}}]{Gao2013}%
  \BibitemOpen
  \bibfield  {author} {\bibinfo {author} {\bibfnamefont {J.}~\bibnamefont
  {Gao}}, \bibinfo {author} {\bibfnamefont {D.~R.}\ \bibnamefont
  {Guildenbecher}}, \bibinfo {author} {\bibfnamefont {P.~L.}\ \bibnamefont
  {Reu}}, \bibinfo {author} {\bibfnamefont {V.}~\bibnamefont {Kulkarni}},
  \bibinfo {author} {\bibfnamefont {P.~E.}\ \bibnamefont {Sojka}},\ and\
  \bibinfo {author} {\bibfnamefont {J.}~\bibnamefont {Chen}},\ }\bibfield
  {title} {\enquote {\bibinfo {title} {Quantitative, three-dimensional
  diagnostics of multiphase drop fragmentation via digital in-line
  holography},}\ }\href@noop {} {\bibfield  {journal} {\bibinfo  {journal}
  {Optics Letters}\ }\textbf {\bibinfo {volume} {38}},\ \bibinfo {pages}
  {1893--1895} (\bibinfo {year} {2013})}\BibitemShut {NoStop}%
\bibitem [{\citenamefont {Radhakrishna}\ \emph {et~al.}(2021)\citenamefont
  {Radhakrishna}, \citenamefont {Shang}, \citenamefont {Yao}, \citenamefont
  {Chen},\ and\ \citenamefont {Sojka}}]{Radhakrishna2021}%
  \BibitemOpen
  \bibfield  {author} {\bibinfo {author} {\bibfnamefont {V.}~\bibnamefont
  {Radhakrishna}}, \bibinfo {author} {\bibfnamefont {W.}~\bibnamefont {Shang}},
  \bibinfo {author} {\bibfnamefont {L.}~\bibnamefont {Yao}}, \bibinfo {author}
  {\bibfnamefont {J.}~\bibnamefont {Chen}},\ and\ \bibinfo {author}
  {\bibfnamefont {P.~E.}\ \bibnamefont {Sojka}},\ }\bibfield  {title} {\enquote
  {\bibinfo {title} {Experimental characterization of secondary atomization at
  high ohnesorge numbers},}\ }\href@noop {} {\bibfield  {journal} {\bibinfo
  {journal} {International Journal of Multiphase Flow}\ }\textbf {\bibinfo
  {volume} {38}},\ \bibinfo {pages} {103591} (\bibinfo {year}
  {2021})}\BibitemShut {NoStop}%
\bibitem [{\citenamefont {Yang}\ \emph {et~al.}(2017)\citenamefont {Yang},
  \citenamefont {Jia}, \citenamefont {Che}, \citenamefont {Sun},\ and\
  \citenamefont {Wang}}]{Yang2017}%
  \BibitemOpen
  \bibfield  {author} {\bibinfo {author} {\bibfnamefont {W.}~\bibnamefont
  {Yang}}, \bibinfo {author} {\bibfnamefont {M.}~\bibnamefont {Jia}}, \bibinfo
  {author} {\bibfnamefont {Z.}~\bibnamefont {Che}}, \bibinfo {author}
  {\bibfnamefont {K.}~\bibnamefont {Sun}},\ and\ \bibinfo {author}
  {\bibfnamefont {T.}~\bibnamefont {Wang}},\ }\bibfield  {title} {\enquote
  {\bibinfo {title} {Transitions of deformation to bag breakup and bag to
  bag-stamen breakup for droplets subjected to a continuous gas flow},}\
  }\href@noop {} {\bibfield  {journal} {\bibinfo  {journal} {International
  Journal of Heat and Mass Transfer}\ }\textbf {\bibinfo {volume} {111}},\
  \bibinfo {pages} {884--894} (\bibinfo {year} {2017})}\BibitemShut {NoStop}%
\bibitem [{\citenamefont {Fang}\ \emph {et~al.}(2022)\citenamefont {Fang},
  \citenamefont {Zhang}, \citenamefont {Jiang}, \citenamefont {Lv},
  \citenamefont {Sun}, \citenamefont {Li}, \citenamefont {Song},\ and\
  \citenamefont {Feng}}]{Fang2022}%
  \BibitemOpen
  \bibfield  {author} {\bibinfo {author} {\bibfnamefont {W.}~\bibnamefont
  {Fang}}, \bibinfo {author} {\bibfnamefont {K.}~\bibnamefont {Zhang}},
  \bibinfo {author} {\bibfnamefont {Q.}~\bibnamefont {Jiang}}, \bibinfo
  {author} {\bibfnamefont {C.}~\bibnamefont {Lv}}, \bibinfo {author}
  {\bibfnamefont {C.}~\bibnamefont {Sun}}, \bibinfo {author} {\bibfnamefont
  {Q.}~\bibnamefont {Li}}, \bibinfo {author} {\bibfnamefont {Y.}~\bibnamefont
  {Song}},\ and\ \bibinfo {author} {\bibfnamefont {X.-Q.}\ \bibnamefont
  {Feng}},\ }\bibfield  {title} {\enquote {\bibinfo {title} {Drop impact
  dynamics on solid surfaces},}\ }\href@noop {} {\bibfield  {journal} {\bibinfo
   {journal} {Applied Physics Letters}\ }\textbf {\bibinfo {volume} {121}},\
  \bibinfo {pages} {210501} (\bibinfo {year} {2022})}\BibitemShut {NoStop}%
\bibitem [{\citenamefont {Savva}\ and\ \citenamefont {Bush}(2009)}]{Savva2009}%
  \BibitemOpen
  \bibfield  {author} {\bibinfo {author} {\bibfnamefont {N.}~\bibnamefont
  {Savva}}\ and\ \bibinfo {author} {\bibfnamefont {J.~W.}\ \bibnamefont
  {Bush}},\ }\bibfield  {title} {\enquote {\bibinfo {title} {Viscous sheet
  retraction},}\ }\href@noop {} {\bibfield  {journal} {\bibinfo  {journal}
  {Journal of Fluid Mechanics}\ }\textbf {\bibinfo {volume} {626}},\ \bibinfo
  {pages} {211--240} (\bibinfo {year} {2009})}\BibitemShut {NoStop}%
\bibitem [{\citenamefont {Lhuissier}\ and\ \citenamefont
  {Villermaux}(2012)}]{Lhuissier2012}%
  \BibitemOpen
  \bibfield  {author} {\bibinfo {author} {\bibfnamefont {H.}~\bibnamefont
  {Lhuissier}}\ and\ \bibinfo {author} {\bibfnamefont {E.}~\bibnamefont
  {Villermaux}},\ }\bibfield  {title} {\enquote {\bibinfo {title} {Bursting
  bubble aerosols},}\ }\href@noop {} {\bibfield  {journal} {\bibinfo  {journal}
  {Journal of Fluid Mechanics}\ }\textbf {\bibinfo {volume} {696}},\ \bibinfo
  {pages} {5--44} (\bibinfo {year} {2012})}\BibitemShut {NoStop}%
\bibitem [{\citenamefont {Culick}(1960)}]{Culick1960}%
  \BibitemOpen
  \bibfield  {author} {\bibinfo {author} {\bibfnamefont {F.~E.}\ \bibnamefont
  {Culick}},\ }\bibfield  {title} {\enquote {\bibinfo {title} {Comments on a
  ruptured soap film},}\ }\href@noop {} {\bibfield  {journal} {\bibinfo
  {journal} {Journal of applied physics}\ }\textbf {\bibinfo {volume} {31}},\
  \bibinfo {pages} {1128--1129} (\bibinfo {year} {1960})}\BibitemShut {NoStop}%
\bibitem [{\citenamefont {Taylor}(1959)}]{Taylor1959}%
  \BibitemOpen
  \bibfield  {author} {\bibinfo {author} {\bibfnamefont {G.~I.}\ \bibnamefont
  {Taylor}},\ }\bibfield  {title} {\enquote {\bibinfo {title} {The dynamics of
  thin sheets of fluid. iii. disintegration of fluid sheets},}\ }\href@noop {}
  {\bibfield  {journal} {\bibinfo  {journal} {Proceedings of the Royal Society
  of London. Series A. Mathematical and Physical Sciences}\ }\textbf {\bibinfo
  {volume} {253}},\ \bibinfo {pages} {313--321} (\bibinfo {year}
  {1959})}\BibitemShut {NoStop}%
\bibitem [{\citenamefont {Pfeiffer}\ \emph {et~al.}(2020)\citenamefont
  {Pfeiffer}, \citenamefont {Zeng}, \citenamefont {Tan},\ and\ \citenamefont
  {Ohl}}]{Pfeiffer2020}%
  \BibitemOpen
  \bibfield  {author} {\bibinfo {author} {\bibfnamefont {P.}~\bibnamefont
  {Pfeiffer}}, \bibinfo {author} {\bibfnamefont {Q.}~\bibnamefont {Zeng}},
  \bibinfo {author} {\bibfnamefont {B.~H.}\ \bibnamefont {Tan}},\ and\ \bibinfo
  {author} {\bibfnamefont {C.-D.}\ \bibnamefont {Ohl}},\ }\bibfield  {title}
  {\enquote {\bibinfo {title} {Merging of soap bubbles and why surfactant
  matters},}\ }\href@noop {} {\bibfield  {journal} {\bibinfo  {journal}
  {Applied Physics Letters}\ }\textbf {\bibinfo {volume} {116}},\ \bibinfo
  {pages} {103702} (\bibinfo {year} {2020})}\BibitemShut {NoStop}%
\bibitem [{\citenamefont {Keller}\ and\ \citenamefont
  {Kolodner}(1954)}]{Keller1954}%
  \BibitemOpen
  \bibfield  {author} {\bibinfo {author} {\bibfnamefont {J.~B.}\ \bibnamefont
  {Keller}}\ and\ \bibinfo {author} {\bibfnamefont {I.}~\bibnamefont
  {Kolodner}},\ }\bibfield  {title} {\enquote {\bibinfo {title} {Instability of
  liquid surfaces and the formation of drops},}\ }\href@noop {} {\bibfield
  {journal} {\bibinfo  {journal} {Journal of Applied Physics}\ }\textbf
  {\bibinfo {volume} {25}},\ \bibinfo {pages} {918--921} (\bibinfo {year}
  {1954})}\BibitemShut {NoStop}%
\bibitem [{\citenamefont {Vledouts}\ \emph {et~al.}(2016)\citenamefont
  {Vledouts}, \citenamefont {Quinard}, \citenamefont {Vandenberghe},\ and\
  \citenamefont {Villermaux}}]{Vledouts2016}%
  \BibitemOpen
  \bibfield  {author} {\bibinfo {author} {\bibfnamefont {A.}~\bibnamefont
  {Vledouts}}, \bibinfo {author} {\bibfnamefont {J.}~\bibnamefont {Quinard}},
  \bibinfo {author} {\bibfnamefont {N.}~\bibnamefont {Vandenberghe}},\ and\
  \bibinfo {author} {\bibfnamefont {E.}~\bibnamefont {Villermaux}},\ }\bibfield
   {title} {\enquote {\bibinfo {title} {Explosive fragmentation of liquid
  shells},}\ }\href@noop {} {\bibfield  {journal} {\bibinfo  {journal} {Journal
  of Fluid Mechanics}\ }\textbf {\bibinfo {volume} {788}},\ \bibinfo {pages}
  {246--273} (\bibinfo {year} {2016})}\BibitemShut {NoStop}%
\bibitem [{\citenamefont {Wang}\ and\ \citenamefont
  {Bourouiba}(2022)}]{Wang2022}%
  \BibitemOpen
  \bibfield  {author} {\bibinfo {author} {\bibfnamefont {Y.}~\bibnamefont
  {Wang}}\ and\ \bibinfo {author} {\bibfnamefont {L.}~\bibnamefont
  {Bourouiba}},\ }\bibfield  {title} {\enquote {\bibinfo {title} {Mass,
  momentum and energy partitioning in unsteady fragmentation},}\ }\href@noop {}
  {\bibfield  {journal} {\bibinfo  {journal} {Journal of Fluid Mechanics}\
  }\textbf {\bibinfo {volume} {935}},\ \bibinfo {pages} {A29} (\bibinfo {year}
  {2022})}\BibitemShut {NoStop}%
\bibitem [{\citenamefont {Jalaal}\ and\ \citenamefont
  {Mehravaran}(2012)}]{Jalaal2012}%
  \BibitemOpen
  \bibfield  {author} {\bibinfo {author} {\bibfnamefont {M.}~\bibnamefont
  {Jalaal}}\ and\ \bibinfo {author} {\bibfnamefont {K.}~\bibnamefont
  {Mehravaran}},\ }\bibfield  {title} {\enquote {\bibinfo {title}
  {Fragmentation of falling liquid droplets in bag breakup mode},}\ }\href@noop
  {} {\bibfield  {journal} {\bibinfo  {journal} {International Journal of
  Multiphase Flow}\ }\textbf {\bibinfo {volume} {47}},\ \bibinfo {pages}
  {115--132} (\bibinfo {year} {2012})}\BibitemShut {NoStop}%
\bibitem [{\citenamefont {Mullagura}\ and\ \citenamefont
  {Dash}(2020)}]{Mullagura2020}%
  \BibitemOpen
  \bibfield  {author} {\bibinfo {author} {\bibfnamefont {H.~N.}\ \bibnamefont
  {Mullagura}}\ and\ \bibinfo {author} {\bibfnamefont {S.}~\bibnamefont
  {Dash}},\ }\bibfield  {title} {\enquote {\bibinfo {title} {Bubble-induced
  rupture of droplets on hydrophobic and lubricant-impregnated surfaces},}\
  }\href@noop {} {\bibfield  {journal} {\bibinfo  {journal} {Langmuir}\
  }\textbf {\bibinfo {volume} {36}},\ \bibinfo {pages} {8858--8864} (\bibinfo
  {year} {2020})}\BibitemShut {NoStop}%
\bibitem [{\citenamefont {Naidu}\ and\ \citenamefont {Dash}(2022)}]{Naidu2022}%
  \BibitemOpen
  \bibfield  {author} {\bibinfo {author} {\bibfnamefont {D.~P.}\ \bibnamefont
  {Naidu}}\ and\ \bibinfo {author} {\bibfnamefont {S.}~\bibnamefont {Dash}},\
  }\bibfield  {title} {\enquote {\bibinfo {title} {Impact dynamics of
  air-in-liquid compound droplets},}\ }\href@noop {} {\bibfield  {journal}
  {\bibinfo  {journal} {Physics of Fluids}\ }\textbf {\bibinfo {volume} {34}},\
  \bibinfo {pages} {073604} (\bibinfo {year} {2022})}\BibitemShut {NoStop}%
\bibitem [{\citenamefont {Vadivukkarasan}, \citenamefont {Dhivyaraja},\ and\
  \citenamefont {Panchagnula}(2020)}]{Vadivukkarasan2020}%
  \BibitemOpen
  \bibfield  {author} {\bibinfo {author} {\bibfnamefont {M.}~\bibnamefont
  {Vadivukkarasan}}, \bibinfo {author} {\bibfnamefont {K.}~\bibnamefont
  {Dhivyaraja}},\ and\ \bibinfo {author} {\bibfnamefont {M.~V.}\ \bibnamefont
  {Panchagnula}},\ }\bibfield  {title} {\enquote {\bibinfo {title} {Breakup
  morphology of expelled respiratory liquid: From the perspective of
  hydrodynamic instabilities},}\ }\href@noop {} {\bibfield  {journal} {\bibinfo
   {journal} {Physics of Fluids}\ }\textbf {\bibinfo {volume} {32}},\ \bibinfo
  {pages} {094101} (\bibinfo {year} {2020})}\BibitemShut {NoStop}%
\bibitem [{\citenamefont {Mehrabian}\ and\ \citenamefont
  {Feng}(2013)}]{Mehrabian2013}%
  \BibitemOpen
  \bibfield  {author} {\bibinfo {author} {\bibfnamefont {H.}~\bibnamefont
  {Mehrabian}}\ and\ \bibinfo {author} {\bibfnamefont {J.~J.}\ \bibnamefont
  {Feng}},\ }\bibfield  {title} {\enquote {\bibinfo {title} {Capillary breakup
  of a liquid torus},}\ }\href@noop {} {\bibfield  {journal} {\bibinfo
  {journal} {Journal of Fluid Mechanics}\ }\textbf {\bibinfo {volume} {717}},\
  \bibinfo {pages} {281--292} (\bibinfo {year} {2013})}\BibitemShut {NoStop}%
\bibitem [{\citenamefont {Pairam}\ and\ \citenamefont
  {Fern{\'a}ndez-Nieves}(2009)}]{Pairam2009}%
  \BibitemOpen
  \bibfield  {author} {\bibinfo {author} {\bibfnamefont {E.}~\bibnamefont
  {Pairam}}\ and\ \bibinfo {author} {\bibfnamefont {A.}~\bibnamefont
  {Fern{\'a}ndez-Nieves}},\ }\bibfield  {title} {\enquote {\bibinfo {title}
  {Generation and stability of toroidal droplets in a viscous liquid},}\
  }\href@noop {} {\bibfield  {journal} {\bibinfo  {journal} {Physical Review
  Letters}\ }\textbf {\bibinfo {volume} {102}},\ \bibinfo {pages} {234501}
  (\bibinfo {year} {2009})}\BibitemShut {NoStop}%
\bibitem [{\citenamefont {Kulkarni}\ \emph {et~al.}(2021)\citenamefont
  {Kulkarni}, \citenamefont {Lolla}, \citenamefont {Tamvada}, \citenamefont
  {Shirdade},\ and\ \citenamefont {Anand}}]{Kulkarni2021}%
  \BibitemOpen
  \bibfield  {author} {\bibinfo {author} {\bibfnamefont {V.}~\bibnamefont
  {Kulkarni}}, \bibinfo {author} {\bibfnamefont {V.~Y.}\ \bibnamefont {Lolla}},
  \bibinfo {author} {\bibfnamefont {S.~R.}\ \bibnamefont {Tamvada}}, \bibinfo
  {author} {\bibfnamefont {N.}~\bibnamefont {Shirdade}},\ and\ \bibinfo
  {author} {\bibfnamefont {S.}~\bibnamefont {Anand}},\ }\bibfield  {title}
  {\enquote {\bibinfo {title} {Coalescence and spreading of drops on liquid
  pools},}\ }\href@noop {} {\bibfield  {journal} {\bibinfo  {journal} {Journal
  of Colloid and Interface Science}\ }\textbf {\bibinfo {volume} {586}},\
  \bibinfo {pages} {257--268} (\bibinfo {year} {2021})}\BibitemShut {NoStop}%
\end{thebibliography}%


\begin{thebibliography}{25}
\providecommand{\natexlab}[1]{#1}
\providecommand{\url}[1]{\texttt{#1}}
\expandafter\ifx\csname urlstyle\endcsname\relax
  \providecommand{\doi}[1]{doi: #1}\else
  \providecommand{\doi}{doi: \begingroup \urlstyle{rm}\Url}\fi

\bibitem[Kulkarni(2013)]{Kulkarni2013}
Varun Kulkarni.
\newblock \emph{An analytical and experimental study of secondary atomization
  of vibrational and bag breakup modes}.
\newblock PhD thesis, Purdue University, 2013.

\bibitem[Kulkarni and Sojka(2014)]{Kulkarni2014}
Varun Kulkarni and Paul~E Sojka.
\newblock Bag breakup of low viscosity drops in the presence of a continuous
  air jet.
\newblock \emph{Physics of Fluids}, 26\penalty0 (7):\penalty0 072103, 2014.

\bibitem[Zhao et~al.(2011{\natexlab{a}})Zhao, Liu, Xu, and Li]{Zhao2011}
Hui Zhao, Hai~Feng Liu, Jian~Liang Xu, and Wei~Feng Li.
\newblock Experimental study of drop size distribution in the bag breakup
  regime.
\newblock \emph{Industrial \& engineering chemistry research}, 50\penalty0
  (16):\penalty0 9767--9773, 2011{\natexlab{a}}.

\bibitem[Jackiw and Ashgriz(2022)]{Jackiw2022}
Isaac~M Jackiw and Nasser Ashgriz.
\newblock Prediction of the droplet size distribution in aerodynamic droplet
  breakup.
\newblock \emph{Journal of Fluid Mechanics}, 940:\penalty0 A17, 2022.

\bibitem[Gao et~al.(2013)Gao, Guildenbecher, Reu, Kulkarni, Sojka, and
  Chen]{Gao2013}
Jian Gao, Daniel~R Guildenbecher, Phillip~L Reu, Varun Kulkarni, Paul~E Sojka,
  and Jun Chen.
\newblock Quantitative, three-dimensional diagnostics of multiphase drop
  fragmentation via digital in-line holography.
\newblock \emph{Optics letters}, 38\penalty0 (11):\penalty0 1893--1895, 2013.

\bibitem[Radhakrishna et~al.(2021)Radhakrishna, Shang, Yao, Chen, and
  Sojka]{Radhakrishna2021}
Vishnu Radhakrishna, Weixiao Shang, Longchao Yao, Jun Chen, and Paul~E Sojka.
\newblock Experimental characterization of secondary atomization at high
  ohnesorge numbers.
\newblock \emph{International Journal of Multiphase Flow}, 138:\penalty0
  103591, 2021.

\bibitem[Guildenbecher et~al.(2017)Guildenbecher, Gao, Chen, and
  Sojka]{Guildenbecher2017}
Daniel~R Guildenbecher, Jian Gao, Jun Chen, and Paul~E Sojka.
\newblock Characterization of drop aerodynamic fragmentation in the bag and
  sheet-thinning regimes by crossed-beam, two-view, digital in-line holography.
\newblock \emph{International journal of multiphase flow}, 94:\penalty0
  107--122, 2017.

\bibitem[Guildenbecher et~al.(2009)Guildenbecher, L{\'o}pez-Rivera, and
  Sojka]{Guildenbecher2009}
DR~Guildenbecher, C~L{\'o}pez-Rivera, and PE~Sojka.
\newblock Secondary atomization.
\newblock \emph{Experiments in Fluids}, 46\penalty0 (3):\penalty0 371--402,
  2009.

\bibitem[Villermaux(2007)]{Villermaux2007}
Emmanuel Villermaux.
\newblock Fragmentation.
\newblock \emph{Annu. Rev. Fluid Mech.}, 39:\penalty0 419--446, 2007.

\bibitem[Villermaux and Bossa(2009)]{Villermaux2009}
Emmanuel Villermaux and Benjamin Bossa.
\newblock Single-drop fragmentation determines size distribution of raindrops.
\newblock \emph{Nature physics}, 5\penalty0 (9):\penalty0 697--702, 2009.

\bibitem[Rayleigh(1882)]{Rayleigh1882}
Rayleigh Rayleigh.
\newblock Investigation of the character of the equilibrium of an
  incompressible heavy fluid of variable density.
\newblock \emph{Proceedings of the London mathematical society}, 1\penalty0
  (1):\penalty0 170--177, 1882.

\bibitem[Taylor(1950)]{Taylor1950}
Geoffrey~Ingram Taylor.
\newblock The instability of liquid surfaces when accelerated in a direction
  perpendicular to their planes. i.
\newblock \emph{Proceedings of the Royal Society of London. Series A.
  Mathematical and Physical Sciences}, 201\penalty0 (1065):\penalty0 192--196,
  1950.

\bibitem[Bellman and Pennington(1954)]{Bellman1954}
Richard Bellman and Ralph~H Pennington.
\newblock Effects of surface tension and viscosity on taylor instability.
\newblock \emph{Quarterly of Applied Mathematics}, 12\penalty0 (2):\penalty0
  151--162, 1954.

\bibitem[Zhao et~al.(2010)Zhao, Liu, Li, and Xu]{Zhao2010}
Hui Zhao, Hai-Feng Liu, Wei-Feng Li, and Jian-Liang Xu.
\newblock Morphological classification of low viscosity drop bag breakup in a
  continuous air jet stream.
\newblock \emph{Physics of Fluids}, 22\penalty0 (11):\penalty0 114103, 2010.

\bibitem[Zhao et~al.(2011{\natexlab{b}})Zhao, Liu, Cao, Li, and Xu]{Zhao2011a}
Hui Zhao, Hai-Feng Liu, Xian-Kui Cao, Wei-Feng Li, and Jian-Liang Xu.
\newblock Breakup characteristics of liquid drops in bag regime by a continuous
  and uniform air jet flow.
\newblock \emph{International Journal of Multiphase Flow}, 37\penalty0
  (5):\penalty0 530--534, 2011{\natexlab{b}}.

\bibitem[Pfeiffer et~al.(2020)Pfeiffer, Zeng, Tan, and Ohl]{Pfeiffer2020}
Patricia Pfeiffer, Qingyun Zeng, Beng~Hau Tan, and Claus-Dieter Ohl.
\newblock Merging of soap bubbles and why surfactant matters.
\newblock \emph{Applied Physics Letters}, 116\penalty0 (10):\penalty0 103702,
  2020.

\bibitem[Opfer et~al.(2014)Opfer, Roisman, Venzmer, Klostermann, and
  Tropea]{Opfer2014}
L~Opfer, IV~Roisman, J~Venzmer, M~Klostermann, and C~Tropea.
\newblock Droplet-air collision dynamics: Evolution of the film thickness.
\newblock \emph{Physical Review E}, 89\penalty0 (1):\penalty0 013023, 2014.

\bibitem[Keller and Kolodner(1954)]{Keller1954}
Joseph~B Keller and Ignace Kolodner.
\newblock Instability of liquid surfaces and the formation of drops.
\newblock \emph{Journal of Applied Physics}, 25\penalty0 (7):\penalty0
  918--921, 1954.

\bibitem[Bremond and Villermaux(2005)]{Bremond2005}
Nicolas Bremond and Emmanuel Villermaux.
\newblock Bursting thin liquid films.
\newblock \emph{Journal of fluid mechanics}, 524:\penalty0 121--130, 2005.

\bibitem[Vledouts et~al.(2016)Vledouts, Quinard, Vandenberghe, and
  Villermaux]{Vledouts2016}
A~Vledouts, J~Quinard, N~Vandenberghe, and E~Villermaux.
\newblock Explosive fragmentation of liquid shells.
\newblock \emph{Journal of Fluid Mechanics}, 788:\penalty0 246--273, 2016.

\bibitem[Yarin and Weiss(1995)]{Yarin1995}
Alexander~L Yarin and Daniel~A Weiss.
\newblock Impact of drops on solid surfaces: self-similar capillary waves, and
  splashing as a new type of kinematic discontinuity.
\newblock \emph{Journal of fluid mechanics}, 283:\penalty0 141--173, 1995.

\bibitem[Wang and Bourouiba(2017)]{Wang2017}
Y~Wang and L~Bourouiba.
\newblock Drop impact on small surfaces: thickness and velocity profiles of the
  expanding sheet in the air.
\newblock \emph{Journal of Fluid Mechanics}, 814:\penalty0 510--534, 2017.

\bibitem[Lhuissier and Villermaux(2012)]{Lhuissier2012}
Henri Lhuissier and Emmanuel Villermaux.
\newblock Bursting bubble aerosols.
\newblock \emph{Journal of Fluid Mechanics}, 696:\penalty0 5--44, 2012.

\bibitem[Wang and Bourouiba(2022)]{Wang2022}
Y~Wang and L~Bourouiba.
\newblock Mass, momentum and energy partitioning in unsteady fragmentation.
\newblock \emph{Journal of Fluid Mechanics}, 935:\penalty0 A29, 2022.

\bibitem[Poulain et~al.(2018)Poulain, Villermaux, and Bourouiba]{Poulain2018}
S~Poulain, E~Villermaux, and L~Bourouiba.
\newblock Ageing and burst of surface bubbles.
\newblock \emph{Journal of fluid mechanics}, 851:\penalty0 636--671, 2018.

\end{thebibliography}
\end{document}


\maketitle
\tableofcontents
\thispagestyle{empty}

\newpage

\section{Experimental determination of drop sizes, validation with existing experiments and theory}
The following discussion elaborates upon on the image processing procedure adopted to determine the rim and film drop sizes, their validation with existing experimental measurements and theoretical arguments to demonstrate their veracity. 
\subsection{Details of image processing to obtain drop sizes} {\label{Section3}}
Our experiments involve capturing videos using Vision Research Phantom v7 high speed digital camera at 4700 fps with an exposure time of 100 $\mu s$ and resolution of 800 $\times$ 600 pixels. A 105 mm focal length lens (Nikon AF Micro Nikkor) was attached to the camera and placed such that the videos were recorded from the front similar to our earlier study. \citep{Kulkarni2013, Kulkarni2014} The breakup and deformation of the drop was illuminated using a 1000 W Xenon arc lamp (Kratos model LH151N) as a light source. In this section we provide details of the video/image processing procedure adopted by us to determine rim, film drops sizes once the videos were captured. The steps listed below use the open source application \href{https://imagej.nih.gov/ij/}{\color{black}{NIH, \underline{ImageJ}}}.
\begin{figure}[htp!]
	\centerline{\includegraphics[scale=1.0]{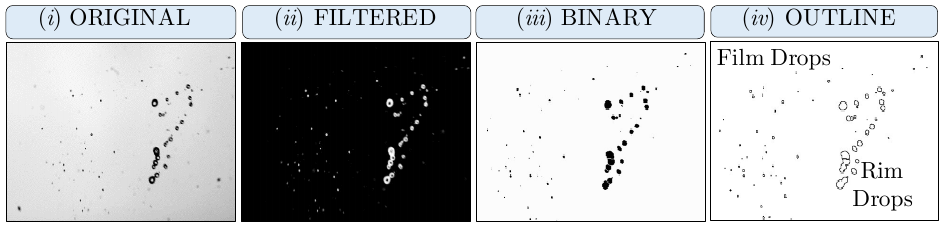}}
   \vspace{0pt}
   \caption{Different stages of image processing (\textit{i}) Raw unprocessed image only corrected for brightness and contrast to give a uniform background. Thereafter, it is converted to one with a white background (\textit{ii}) Filtering to identify the edges (\textit{iii}) Binarization of the image  (\textit{iv}) Segmentation of the binary images to produce outlines of all drops.}
   \label{FigS2}
   \vspace{-10pt}
\end{figure} 
\begin{enumerate}[label=\textup{(}\roman*\textup{)},font=\itshape,leftmargin=0.24in]
\item \textbf{Homogenizing background of raw image}: The recorded videos are first imported in \href{https://imagej.nih.gov/ij/}{\color{black}{NIH, \underline{ImageJ}}} to extract the requisite details. However, in a typically recorded video (8-bit grayscale) the background illumination is not even. To eliminate measurement errors which may be so introduced, we begin by appropriately homogenizing the background lighting by adjusting the brightness and contrast of the video. The background of one such image after this operation on the image sequence generated from the video is shown in Fig. \ref{FigS2}\textcolor{black}{(\textit{i})}. Next, we take a bare background image which does not contain any drops and divide it from the brightness/contrast adjusted image sequence previously obtained. The resulting image sequence has a white background and a clearly visible deformed drop or/and drop fragments in the foreground. 
\item \textbf{Filtering}: The white background image is then processed to determine the edges. The Sobel operator based filter frequently employed in edge detection algorithms is used to identify the boundaries of the drop fragment features in the image. A result of this filtering operation is shown in Fig. \ref{FigS2}\textcolor{black}{(\textit{ii})}. We assume that the gradient operation to identify the periphery of the drop yields nearly isotropic values around the drop.
\item \textbf{Binary}: Once a filtered image is obtained we invert its colors. The grayscale image so obtained is converted into a binary format by defining a gray-scale cutoff point, below which the fragments become black and above which they become white. Appropriate threshold values are determined based on most of the fragment drops being distinctly identified. The final result of this operation is shown in Fig. \ref{FigS2}\textcolor{black}{(\textit{iii})}.
\item \textbf{Outline}: In the last step, we segment the image obtained from the above procedure to demarcate its boundary (see Fig. \ref{FigS2}\textcolor{black}{(\textit{iv})}). The drops which are sufficiently in-focus all around their perimeter as identified in the previous steps are selected and their projected area $A$ is measured on the image. Note that threshold at half value on the gray scale was performed which locates a droplet boundary that is less dependent on defocus. Finally, we compute the diameter ($D$) of the fragments from their projected area, $A$ as $D = \sqrt{4A/ \pi}$. Since the area could potentially vary in small increments we get a finer resolution above $65\;\mu m$.
\end{enumerate}
The drop size measurements obtained after the above steps are repeated for 3 different trials for a given $We$ and liquid. From this data we are able to easily isolate the diameters of the rim drops from the film drops since they are small in number and large in size compared to the film drops. The minimum drop size we measured was $65\;\mu m$ which was computed from a measured area of $4225\;\mu m^2$. Finally, the average of these measurements for a particular $We$ and liquid is calculated and reported as $\langle D_{rim}\rangle$ and $\langle D_{film}\rangle$ in our paper which is then used for deriving their respective scaling with $We$. It is worth mentioning that the equivalent diameter of aspherical drops is considered as these eventually assume a spherical shape consistent with previous such measurements \citep{Zhao2011}.

\subsection{Comparison with previous experimental results}
The size of the fragmented drops is heavily dependent on the initial size of the drops and the chosen $We$. Larger drops produce bigger fragment/ drops which decrease in size with increasing $We$. Here we compare our measurements with selected literature \cite{Jackiw2022, Zhao2011, Gao2013, Radhakrishna2021, Guildenbecher2017} on the topic. We also emphasize that besides the initial drop size even the method used to fragment the initial drop affects the eventual drops sizes. For example, shock tube results may differ from those obtained by fragmentation due to impact with an air stream since the former subjects the drop to rapid deformation while the latter is more gradual. The criterion for the equivalence between the two methods can be found in the review paper by Guildenbecher et al.(2009)\citep{Guildenbecher2009}. 
Table \ref{TableR1} shows experimental data collected from various sources.
\begin{table}[htp!]
\centering
{
\begin{tabular}{ccccc}
\Xhline{1\arrayrulewidth}
\multirow{1}{*}{Authors Name} & \multirow{1}{*}{Film drop sizes}\qquad  & \multirow{1}{*}{Rim drop sizes} \qquad& \multirow{1}{*}{Initial drop size}\qquad  &\multirow{1}{*}{$We$} \\
\Xhline{1\arrayrulewidth}
Zhao et al.(2011) \citep{Zhao2011} \qquad & $\geq 120$\qquad & 800 $-$ 2400\qquad& 4100 $\pm$ 5900 & 9$-$23 \\
Gao et al. (2013)\citep{Gao2013} \qquad & $\geq 100$\qquad& n.a.\qquad & 2580 & 11 \\
Guildenbecher et al. (2017) \citep{Guildenbecher2017} \qquad & $\geq 65$\qquad & 100 $-$ 200\qquad& 2300 & 13 \\
Radhakrishna et al. (2021) \citep{Radhakrishna2021} \qquad & $\geq 75$\qquad & 300 $-$ 400\qquad& 3000 & 14 \\
Jackiw and Ashgriz (2022) \citep{Jackiw2022} \qquad & $\geq 85$\qquad& 300 $-$ 500\qquad& $1900 \pm 200$ & 7$-$15\\
Present Study \qquad & $\geq 65$\qquad & 350 $-$ 600\qquad& 2100$-$2600 & 12$-$20 \\
\Xhline{1\arrayrulewidth}
\end{tabular}
}
\caption{Measured film and rim drops (in $\mu m$) with associated initial drop diameter (in $\mu m$) at a given $We$ showing that our measurement of the film drops above $65 \mu m$ indeed fall in the range of previously reported measurements of this quantity. Data not available indicated as n.a.}
\label{TableR1}
\end{table}
The latest work by Jackiw and Ashgriz (2022) \citep{Jackiw2022} uses shadowgraphy while Gao et al.(2013)\citep{Gao2013} use digital holography validated by Phase Doppler Anemometry (PDA) measurements to measure drop sizes. Radhakrishna et al.(2021)\citep{Radhakrishna2021} and Guildenbecher et al.(2017)\citep{Guildenbecher2017} too use digital holography to measure fragment sizes with the latter being able to measure drops as small as $27\;\mu m$. All of them report fragmented drop sizes in excess of $65 \mu m$ and therefore our chosen resolution is sufficient.

\subsection{Theoretical proof that most fragment volume lies in drops $\geq 65\;\mu m$}
\begin{figure}[htp!]
	\centerline{\includegraphics[scale=0.75]{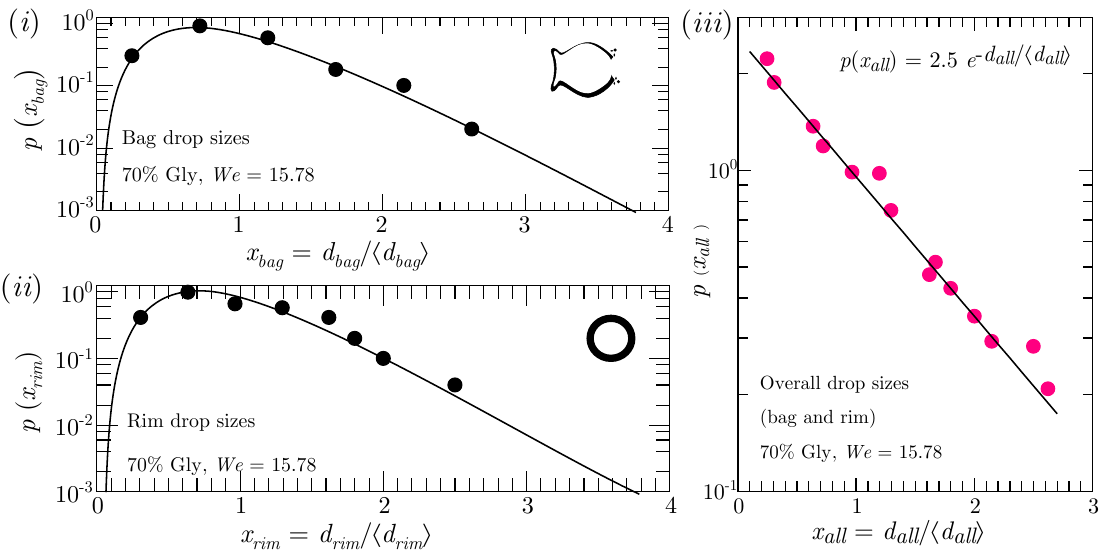}}
   \vspace{0pt}
   \caption{Drop size distribution fitted with gamma function ($n \approx 3.7$) given by eqn \ref{oneGD} for (\textit{i}) bag drops and, (\textit{ii}) rim drops. (\textit{iii}) Combined rim and bag drop size distribution displaying an exponential distribution. Test conditions correspond to 70\% glycerine-water solution at $We = 15.78$. The same trend is seen for all tested liquids and $We$. All plots are drawn on a semi-log scale.}
   \label{FigS21}
   \vspace{-10pt}
\end{figure} 
Fragmentation of liquids via formation of ligaments has been related to aggregation scenario which result in drop sizes which can be described by gamma distribution\citep{Villermaux2007}. The two parameter form of this distribution is given by,\citep{Jackiw2022}
\begin{equation}\label{twoGD}
\mathcal{F}\left(x = d/s\right) = \dfrac{x^{\alpha-1}e^{-x/\beta}}{\beta^{\alpha}\Gamma\left(\alpha\right)}
\end{equation}
In the above, $\alpha$ is the shape factor, $\beta$ is the scale factor and $s$ is a suitable constant. Choosing $s$ such that, $s = \langle d \rangle = D_{10}$ (number mean diameter) implies, $\langle x \rangle = 1$ and $\beta = \alpha^{-1}$ since from the definition\citep{Jackiw2022} of $\alpha$ and $\beta$ we can write, $\alpha\beta = \langle x\rangle$. This reduces eqn \ref{twoGD} to a single parameter distribution with a mean, $\langle d \rangle$ and written as,
\begin{equation}\label{oneGD}
\mathcal{P}\left(x = d/\langle d \rangle\right) = \dfrac{n^n}{\Gamma\left(n\right)}x^{n-1}e^{-nx}
\end{equation}
where, $\alpha$ has been replaced $n$ to facilitate comparison with the commonly used form.\citep{Villermaux2007, Zhao2011}

In our case both the rim and bag drops are produced by a ligament mediated mechanism (details of which are provided in the main manuscript) and therefore amenable to being fitted by eqn \ref{oneGD} as shown in Fig. \ref{FigS21}\textcolor{black}{(\textit{i})} and Fig. \ref{FigS21}\textcolor{black}{(\textit{ii})}. The combined distribution represented as a linear superposition of these two distributions is exponential as seen in Fig. \ref{FigS21}\textcolor{black}{(\textit{iii})} and is established experimentally and theoretically in earlier works \citep{Zhao2011, Villermaux2009, Villermaux2007} on bag breakup and fragmentation.

To prove that drops of sizes $< 65\;\mu m$ contain very less volume of the original drop, we consider the cumulative gamma distribution of the volume\citep{Zhao2011}  of all (rim \& film) drops denoted by the subscript, ``\textit{all}''. For our example case in Fig. \ref{FigS21}\textcolor{black}{(\textit{iii})} we obtain a fit expressed mathematically as, $\mathcal{P}\left(x_{all} = d_{all}/\langle d_{all} \rangle\right) = 2.5 e^{- d_{all}/\langle d_{all} \rangle}$ from which we may write the following cumulative distribution,
\begin{equation}\label{cpdf}
\mathcal{C}\left(x_{all} = d_{all}/\langle d_{all} \rangle\right) = \dfrac{\int\limits_0^{x_{all}} x_{all}^3 e^{-x_{all}} \,\mathrm{d}x_{all}}{\int\limits_0^{\infty} x_{all}^3 e^{-x_{all}} \,\mathrm{d}x_{all}} = 1 - e^{-x_{all}} \left(1 + x_{all} + \frac{x_{all}^2}{2!} + \frac{{x_{all}}^3}{3!}\right)
\end{equation}
In the above, $\mathcal{C}$ is the fraction of the total initial drop volume contained in drops of diameter less than $d_{all}$. For the test condition shown in Fig. \ref{FigS21}, $d_{all} = 65\;\mu m$ for $\langle d_{all} \rangle = D_{10} = 385\;\mu m$ we obtain using eqn \ref{cpdf}, $\mathcal{C}\left(x_{all} = 0.17\right) = 1.29\%$. This means only 1.29\% of the total volume is contained in drops of diameter $< 65\;\mu m$ for this condition. At other $We$ and for other liquids we see this percentage rise only to 3\% hence justifying our current measurement range in excess of $65\;\mu m$.

\section{Details of derivation of equation (5) in the manuscript: $\cfrac{{{{\overline{\lambda}}_{bag}^2}}}{{\overline{\lambda}\;}_{rim}^4\;{\overline{\lambda}}_{film}} \sim We^2$} {\label{SectionS1}}
When a vertically falling drop is impacted by cross stream of air it accelerates rapidly in radial and streamwise directions. Such accelerations are known to induce Rayleigh-Taylor instability since the heavier liquid (water) tries to accelerate into a lighter fluid (air). However, unlike the original case of superposed fluids investigated by Rayleigh\citep{Rayleigh1882} and Taylor\citep{Taylor1950} surface tension plays a significant role here. Accounting for it \citep{Bellman1954} also enables selection of the most amplified or maximum unstable wavelength \citep{Zhao2010,Zhao2011a, Kulkarni2013, Pfeiffer2020} and whose expression in dimensionless form (scaled by the initial drop diameter, $D_0$) can be written as,
\begin{equation}\label{eqn1}
\dfrac{\lambda}{D_0} \sim \sqrt{\dfrac{\sigma}{\rho_l \xi D_0^2}}
\end{equation}
where, $\lambda$ is the most amplified or maximum unstable wavelength [in m], $\rho_l$ is the density of the liquid drop [in kg$\cdot$m\textsuperscript{-3}], $\sigma_l$ is the surface tension of the liquid drop [in N$\cdot$m\textsuperscript{-1}] and $\xi$ acceleration experienced by the drop [in m$\cdot$s\textsuperscript{-2}]. 
To evaluate the corresponding expressions for eqn \ref{eqn1} for the rim and the bag we need to determine the appropriate acceleration, $\xi$ which is not known \textit{apriori}. This translates to evaluating the acceleration in the deforming liquid drop due to air impact. To do so we simply equate the dynamic pressure of the oncoming air stream, $\rho_a U_a^2$, where $\rho_a$ is the density of air [in kg$\cdot$m\textsuperscript{-3}] and $U_a$ is the velocity of the air stream [in m$\cdot$s\textsuperscript{-1}] to the stress induced in the drop manifesting as its inertia, $\rho_l U_l^2$ where, $U_l$ is the radial velocity of the expanding liquid drop [in m$\cdot$s\textsuperscript{-1}] which is also the velocity with which the drop gets squeezed in the streamwise direction. Therefore, we have \citep{Opfer2014},
\begin{equation}\label{eqn2}
U_l \approx \sqrt{\dfrac{\rho_a}{\rho_l}}U_a
\end{equation}
\subsection{Expression for ${\overline{\lambda}}_{rim}$, eqn (1) in the manuscript}{\label{rim}}
We can compute the acceleration in the rim by recognizing the equivalence of two expressions for inertia per unit area (in the radial, \textit{r} direction), the first being, $\rho_l h_{rim}\xi_{rim}$ and the second being (as stated above), $\rho_l U_l^2$. This simplifies to, $\xi_{rim} \sim U_l^2/ h_{rim}$ which can be substituted in the scaling expression, eqn \ref{eqn1} for most unstable wavelength for the rim giving rise to the following,
\begin{equation}\label{eqn3}
\dfrac{\lambda_{rim}}{D_0} \sim \sqrt{\frac{h_{rim}}{D_0}}\sqrt{\frac{\sigma}{\rho_l U_l^2 D_0}}
\end{equation}
Using eqn \ref{eqn2} and replacing $\lambda/ D_0$ by its more concise form, $\overline{\lambda}$ we obtain,
\begin{equation}\label{eqn4}
{\overline{\lambda}}_{rim} \sim \sqrt{\frac{h_{rim}}{D_0}}{We^{-1/2}}
\end{equation}
where, $We = \dfrac{\rho_a U_a^2 D_0}{\sigma}$ and eqn \ref{eqn4} being identical to eqn (1) in the manuscript.
\subsection{Expression for ${\overline{\lambda}}_{bag}$, eqn (2) in the manuscript}{\label{Section2}}
Following the same reasoning as Section \ref{rim} with the difference that we now compute inertia per unit area in the streamwise (\textit{z}) direction. To move forward, we choose the length scale, $D_{bag}$, which is the cross stream dimension of the disc-like deformed drop just before the Rayleigh-Taylor instability leading to the formation of the bag becomes apparent on its surface. The corresponding acceleration hence assumes the form, $\xi_{bag} \sim U_l^2/ D_{bag}$ which can be substituted in eqn \ref{eqn1}. Repeating the same algebra as before we arrive at the scaling relation for most unstable wavelength for the disc-like deformed drop as written below.
\begin{equation}\label{eqn5}
{\overline{\lambda}}_{bag} \sim \sqrt{\frac{D_{bag}}{D_0}}{We^{-1/2}}
\end{equation}
The above expression reproduces eqn (2) in the manuscript exactly. Note that $D_{bag}$ is entirely different from $D_{max}$ which is the maximum cross stream dimension and attained when the drop expands further radially.
\subsection{Expression for ${\overline{\lambda}}_{film}$, eqn (3) in the manuscript}{\label{film}}
Once the bag forms, the curved liquid film so formed expands rapidly, accelerating at the rate \citep{Villermaux2009}, $\xi_{film} \left(\sim U_l^2/D_0\right)$ in the radial direction (normal to the curved hemispherical bag as shown by blue arrows in Fig. 1 (\textit{a})-(\textit{iv})) . The thin central film is thus subjected to Rayleigh-Taylor instability too, much like the rim and disc-like deformed drop but with the crucial difference that waves develop on both interfaces of the film and their wavelength, $\lambda_{film} >> h_{film}$ where, $h_{film}$ is the thickness of the curved liquid film just before it ruptures and known to remain fairly constant from its inception \citep{Villermaux2009}. The expression for most unstable or amplified wavelength in such a case was first derived by Keller and Kolodner(1954) \citep{Keller1954, Bremond2005, Vledouts2016} and expressed in dimensionless form as,
\begin{equation}\label{eqn6}
\dfrac{\lambda_{film}}{D_0} \sim \dfrac{\sigma}{\rho_l \xi_{film}h_{film}}\left(\dfrac{1}{D_0}\right)
\end{equation}
Substituting for $\xi_{film}$ and using the definition of $We$ we may write the following,
\begin{equation}\label{eqn7}
\overline{\lambda}_{film}  \sim \left(\dfrac{h_{film}}{D_0}\right)^{-1} We^{-1}
\end{equation}
Finally, noting that mass (or volume) is conserved between the toroidal rim and the thin central film at all $We$ considered in our work and reasons for which can be found in discussion of Fig. 2 in the main manuscript and Section \ref{SectionS2} in this supplementary material we write in dimensionless form (scaled by $D_0$ and represented by overline), 
\begin{equation}\label{eqn8}
\overline{h}_{rim} \sim \sqrt{\overline{D}_{bag}\;\;\overline{h}_{film}}
\end{equation}
\subsection{Combining ${\overline{\lambda}}_{film}$, ${\overline{\lambda}}_{rim}$ and ${\overline{\lambda}}_{bag}$ to obtain eqn (5) in the manuscript}
\FloatBarrier
\begin{figure}[htp!]
   \vspace{-5pt}
	\centerline{\includegraphics[scale=0.45]{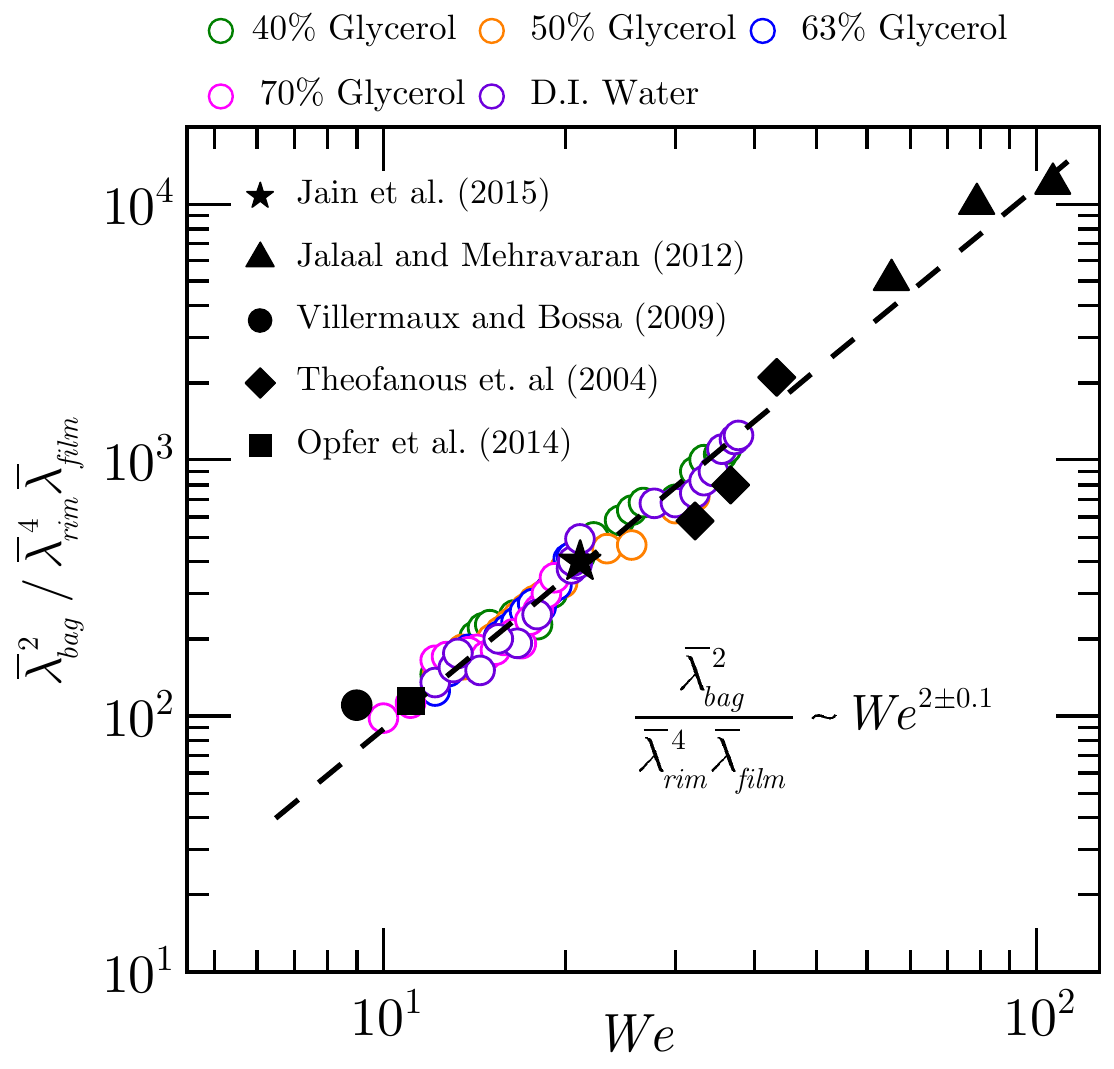}}
   \vspace{0pt}
   \caption{Comparison of our scaling with five previous works with open symbols representing our data. The numerical work by Jalaal and Mehravan (2012) is of particular significance as they use low viscosity drops and observe bag breakup at $We$ as large as 106, which provides some evidence that our scaling works for data spanning 2 decades with $R^2 \approx 0.97$}
   \label{FigS0}
   \vspace{-10pt}
\end{figure} 
Rearranging equations \ref{eqn4}, \ref{eqn5} and \ref{eqn7} derived above such that $\overline{h}_{rim}$, $\overline{D}_{bag}$ and $\overline{h}_{film}$ in the above scaling are expressed in terms wavelengths, $\overline{\lambda}_{rim}$, $\overline{\lambda}_{bag}$ and $\overline{\lambda}_{film}$ and $We$ we write the following,
\begin{align}
\overline{h}_{rim}  & \sim {\overline{\lambda}}_{rim}^4 We                \label{eqn9}   \\
\overline{D}_{bag}  & \sim {\overline{\lambda}}_{bag}^2 We 			       \label{eqn10}   \\
\overline{h}_{film}   & \sim {\overline{\lambda}}_{film}^{-1} We^{-1}   \label{eqn11}  
\end{align}
Finally, substituting eqn \ref{eqn9}, \ref{eqn10} and \ref{eqn11} in eqn \ref{eqn8} we obtain the scaling relation given in eqn (5) of the manuscript (also plotted using open symbols in \ref{FigS0}), 
\begin{equation}\label{eqn12}
\frac{{{{\overline{\lambda}}_{bag}^2}}}{{\overline{\lambda}\;}_{rim}^4\;{\overline{\lambda}}_{film}} \sim We^2
\end{equation}

Next we provide details on how to experimentally measure $\lambda_{rim}$, $\lambda_{film}$ and $\lambda_{bag}$ below,
\begin{itemize}
\item \textbf{For bag} we measure the wavelength ($\lambda_{bag}$) the moment we see an apparent formation of thin central disk bounded by a toroidal rim after the disc-like cylindrical structure ceases to exist (see Fig. 1 (\textit{a})-(\textit{iii})) in the manuscript . This value is also almost equivalent to the inner diameter of the bag film that is later formed and may also be estimated reasonably from there.
\item \textbf{For rim}, the wavelength ($\lambda_{rim}$) is seen when waves form on the surface of the rim circumferentially (along the $\theta$ direction of Fig. 1 (\textit{a})-(\textit{iii}) in the manuscript). However, this may not be very easy to identify in its initial stages especially since we view it from the side. So, alternatively we wait for the instability to grow in amplitude and measure the wavelength when the bag bursts and only the intact rim remains as shown in Fig. 1 (\textit{a})-(\textit{vi}) in the main manuscript. This wavelength does not differ much from its value at its incipience.
\item \textbf{For curved thin film} of the bag we measure the wavelength ($\lambda_{film}$) along the curvature of the bag film in the frame before it bursts (see Fig. 1 (\textit{a})-(\textit{v})).
\end{itemize}
To conclude we also show a comparison between our scaling relation in \ref{eqn12} with existing data from literature (see Fig. \ref{FigS0}) to further validate our result. 

\section{$We$ independent conservation of mass between rim, central film and disc-like deformed drop} \label{SectionS2}
\begin{figure}[htp!]
	\centerline{\includegraphics[scale=0.8]{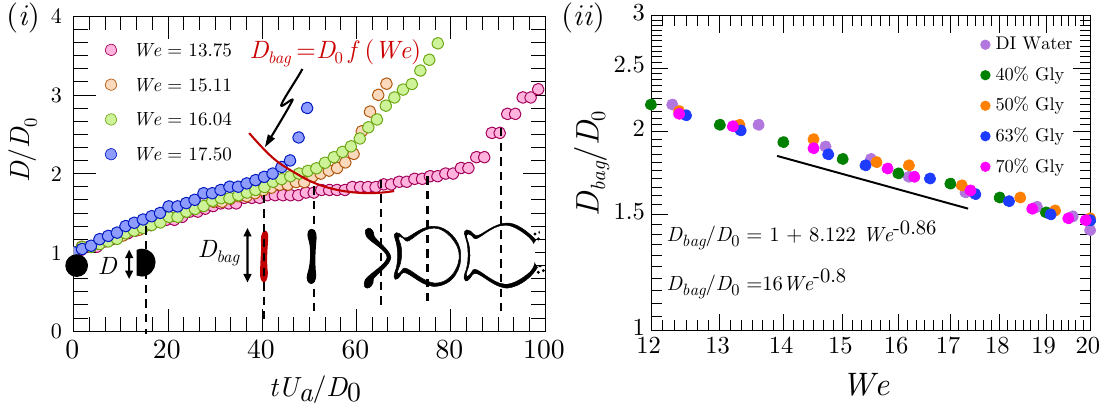}}
   \vspace{0pt}
   \caption{(\textit{i}) Experimental dimensionless cross stream diameter. $D/D_0$ at various times scaled by $D_0/U_a$, where $D_0$ is the initial undeformed drop diameter and $U_a$ is the cross stream air jet velocity. The sketches overlaying the dotted lines show the shape of the deformed drop at that instance of time. The maroon curved line connects all points which correspond to $D_{bag}/D_0$ (\textit{ii}) Scaling for $D_{bag}/D_0$ with $We$ showing an inverse ($We^{-1}$) dependence. Test conditions correspond to $We = 13.75$ and 50\% glycerine-water.}
   \label{FigS1}
   \vspace{0pt}
\end{figure} 
The development of two independent liquid segments namely, the rim and a thin central sheet are characteristic of thinning liquid masses \citep{Yarin1995, Wang2017} which is a consequence of a spatially varying flow profile and surface tension acting on the periphery of the deforming liquid. In our case, the thin central film begins to form once the pancake-like cylindrical deformed drop is influenced by the Rayleigh-Taylor instability \citep{Zhao2010, Zhao2011a} in the streamwise direction (see Fig. \ref{FigS1}(\textit{i}) at $tU_a/D_ 0 = 40$). This thin central film is bounded by a thicker rim both of which thin simultaneously as the drop moves further downstream in the streamwise direction. In this section we show that this thinning is independent of the $We$. 
To prove this we consider the rate of change of rim volume, $\forall_{rim}$ as given below,
\begin{equation}\label{eqn13}
\dfrac{\textrm{d}\forall_{rim}}{\textrm{d}t} = Q_{in}
\end{equation}
Due to equivalence of drop impact/collision phenomena on solids and in air (like our case) as stated in earlier works \citep{Opfer2014} we can write\citep{Yarin1995, Wang2017} the expression for the volume flow rate entering the rim ($Q_{in}$) as,
\begin{equation}\label{eqn14}
Q_{in} = r_{film} h_{film} (u_{film} - u_{rim})
\end{equation}
In the above, $u_{film} = u_{film}\left(r,t\right)$ is the velocity at any point in the thin central film, $u_{rim} = u_{rim}\left(r,t\right)$ is the velocity of the rim and, $h_{film} = h_{film}\left(r,t\right)$ is the thickness of the thin central film. Further, from prior works \citep{Yarin1995, Wang2017} the velocity of the expanding central thin liquid sheet at any time $t$ and radial location, $r_{film}$ is found to be, $u_{film} = h_{film}r_{film}/t$ (details of which can be found in the referenced papers). Substituting for $u_{film}$ in eqn \ref{eqn14} we obtain, $Q_{in} = h_{film}r_{film}^2/t$ assuming $u_{film} >> u_{rim}$. To make progress we take note of the fact that $\forall_{rim}+ \forall_{film} = \frac{\pi}{6}D_0^3$ at all times where, $\forall_{film}$ is the volume contained in the approximately cylindrical thin central film and equals $h_{film}r_{film}^2$ with $\frac{\pi}{6}D_0^3$ representing the original drop volume. We now can recast eqn \ref{eqn13} given this information as follows,
\begin{equation}\label{eqn15}
-\dfrac{\textrm{d}\forall_{film}}{\textrm{d}t} = \dfrac{\forall_{film}}{t}
\end{equation}
We can integrate eqn \ref{eqn15} from $\forall_{film} = 1$ to $\forall_{film}$ on the left hand side and $t = t_0$ to $t + t_0$ on the right hand side where, $t_0$ is the time when the deformed drop assumes a disc-like cylindrical shape with a cross stream length of $D_{bag}$ also referred in Section \ref{Section2} and shown in Fig. \ref{FigS1}\textcolor{black}{(\textit{i})}. On the other hand, $t$ is the measured time after $t_0$. The following expressions for $\forall_{rim}$, $\forall_{film}$ and $\forall_{rim}/\forall_{film}$ may therefore be deduced.
\begin{align}
\forall_{rim}  =   \dfrac{t}{t_0 + t}               									\label{eqn16}   \\
\forall_{film} =   \dfrac{t_0}{t_0 + t}            									\label{eqn17}   \\
\dfrac{\forall_{rim}}{\forall_{film}} =   \dfrac{t}{t_0}            \label{eqn18}   
\end{align}
Eqns \ref{eqn16}, \ref{eqn17} and \ref{eqn18} clearly demonstrate that $\forall_{rim}$, $\forall_{film}$ and the ratio $\forall_{rim}/\forall_{film}$ are independent of $We$. Our experiments show that the deformed drop discernibly separates into a film and rim at $t = 0.5t_0$ with $t_0 \approx 3$ to $5\;ms$ (also see Fig. 1 in the main manuscript). With these considerations, $\forall_{rim}/\forall_{film} \approx 0.5$ as confirmed by our experiments (see Fig. 2 (inset) in the main manuscript).

As a further demonstration of the validity of our result we calculate the scaling for $\overline{D}_{bag}$ from eqn \ref{eqn8} knowing that $\overline{h}_{film}$ and $\overline{h}_{rim}$ both vary as $We^{-1}$ (see Section \ref{Section3} and main manuscript). Doing so, we infer that $\overline{D}_{bag} \sim We^{-1}$ which is exactly confirmed by our experimental data in Fig. \ref{FigS1}\textcolor{black}{(\textit{ii})}. It is known that the Rayleigh-Taylor instability occurs in a time \citep{Lhuissier2012} which scales as, $\left(\sigma/\rho_l\xi^3\right)^{\frac{1}{4}}$. Choosing the time scale\citep{Kulkarni2014}, $D_0/U_a\sqrt{\rho_l/\rho_a}$ and acceleration, $\xi \sim U_l^2/D_0$ we obtain the dimensionless Rayleigh-Taylor instability time as, $\overline{\tau} \sim We^{-0.25}$ in addition to using the simplification, $\rho_a U_a^2 = \rho_l U_l^2$ as required. Moving further, we balance the inertia per unit area of the radially (in $r$ direction) expanding drop, $\rho_l \dot{r}^2$ with the capillary/surface tension pressure, $\sigma/D_0$ acting on it, which upon choosing the velocity scale as $U_l$ gives rise to the scaling, $\overline{\dot{r}} \sim We^{-0.5}$. Multiplying $\overline{\dot{r}}$ with $\overline{\tau}$ as obtained in the preceding expressions we get, $\overline{D}_{bag} = \overline{\dot{r}} \overline{\tau} \sim We^{-0.75}$ which is indeed very close to our experimental scaling exponent of $-0.8$ and the one demanded by our mass conservation $-1$.

To conclude we state that such $We$ independent mass conservation of rim and film has also been reported for drop impact on solids\citep{Wang2022} which our work extends to drop-on-air impacts.

\section{Rolling rim image details for Fig 3 and details of derivation leading to scaling for $\langle \overline{D}_{rim}\rangle$ and $\langle \overline{D}_{film}\rangle$} {\label{Section3}}
Curved liquid films are known to burst by initiation of a hole.\citep{Lhuissier2012,Poulain2018} For a drop undergoing bag breakup a similar curved liquid film presents itself and therefore it is natural to expect a congruent bursting process. In this section we show that the mechanism which governs bursting of bubbles can be extended to bursting of a bag, providing details on film drop production by such a process. 
\begin{figure}[htp!]
	\centerline{\includegraphics[scale=0.8]{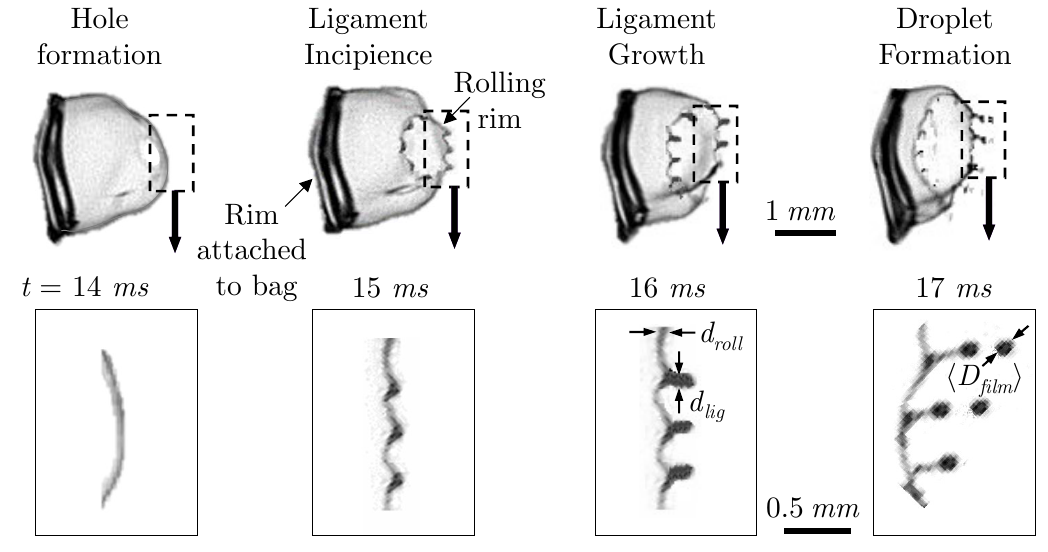}}
   \vspace{0pt}
   \caption{(a) Incipience of a hole, (b) formation of ligaments, (c) elongation of ligaments and, (d) drop generation from ligaments. Test conditions, 50\% glycerine-water, $We \approx 16$}
   \label{FigS3}
   \vspace{-10pt}
\end{figure} 
In Fig. \ref{FigS3} \textcolor{black}{\textit{t} = 14, 15, 16 and 17 \textit{ms}} we show images of the intermediate steps once the hole is formed leading up to the bursting of the curved liquid film. A clear formation of rim is seen which develops once the hole forms in Fig. \ref{FigS3} \textcolor{black}{\textit{t} = 14 \textit{ms}}. As it recedes it is acted upon by surface tension giving rise to a rolling toroidal rim. Such rims, as discussed in the main manuscript are ubiquitous to drop impact phenomena, albeit here our situation is inverted, wherein we have a hole expanding in continuous liquid medium. The rolling toroidal rim is unstable and ejects drops periodically. It is interesting to note that such a rim is fed constantly by the curved liquid film and therefore never depletes in mass until it collides with the rim that bounds the bag. The foregoing argument means that it is reasonable to expect the diameter of rolling rim ($d_{roll}$), the ligaments ($d_{lig}$) and the film drops($D_{film}$) produced to be roughly the same. 

For our next step we refer the schematic used in \textcolor{black}{Fig. 3 (\textit{b})-(\textit{ii})} of the main manuscript to write the volume per unit width in rolling cylindrical rim (assumed to be planar as stated in the manuscript), $\frac{\pi}{4} d_{roll}^2$. This volume is continually fed by the planar sheet (per unit width) of thickness $h_{film}$ at a rate $U_{film}$ and in time, $\tau_{roll}$ corresponding to one time period between successive drop ejections (from Fig. \ref{FigS3} {\textcolor{black}{\textit{t} = 14 to 17 \textit{ms}}}. Therefore, we can write an expression similar to that developed by Lhuissier and Villermaux(2012)\citep{Lhuissier2012} as given below,
\begin{equation}\label{rolldia}
\langle D_{film} \rangle = d_{lig} = d_{roll} = \sqrt{\left(\dfrac{4}{\pi}\right) h_{film} U_{film} \tau_{roll}}
\end{equation}
Replacing, $\tau_{roll}$ by $h_{film}/U_{roll}$ since the film is stretched due to air flow and therefore needs to be separately calculated. Thicker films (with higher $h_{film}$) consequently have larger time between drop ejections. With the substitution for $\tau_{roll}$ eqn \ref{rolldia} now transforms to,
\begin{equation}\label{rolldia}
\langle D_{film} \rangle = \sqrt{\left(\dfrac{4}{\pi}\right) h_{film}^2 \dfrac{U_{film}}{U_{roll}}}
\end{equation}
In view of eqn \ref{rolldia}, to obtain closure we need to determine $U_{film}/U_{roll}$. To do so, we need to evaluate, $\dot{\mathcal{E}}_{film}$ which is the energy per unit time and per unit area assuming unit width of the moving film of depth $h_{film}$. This can be done by considering the force experienced by the moving film due to the dynamic air pressure acting initially on the drop which during the film expansion scales as, $(D_0^2 \rho_l U_{l}^2)$ and since it induces a velocity $U_{film}$ the energy per unit time carried by it is, $(D_0^2 \rho_l U_{l}^2)U_{film}$. The power (energy per unit time) so generated is delivered across a cross sectional area, $h_{film}\cdot1$ of the film assuming unit width which implies, $\dot{\mathcal{E}}_{film} \sim (D_0^2 \rho_l U_{l}^2)U_{film}/h_{film}$. The rolling rim absorbs this energy to maintain its diameter, $d_{roll}$ at all times (which depletes due to drop ejection) moving at $U_{roll}$ corresponding to an energy per unit time given by, $\sigma d_{roll} U_{roll}$. By considering the curved cylindrical surface area of the rolling rim per unit width, $d_{roll}\cdot1$ we can write the surface energy per unit time per unit area required to maintain the diameter of the rim at $d_{roll}$ as, $\dot{\mathcal{E}}_{roll} \sim \sigma d_{roll} U_{roll}/d_{roll}$. Equating $\dot{\mathcal{E}}_{film}$ to $\dot{\mathcal{E}}_{roll}$ results in,
\begin{equation}\label{UfilmUroll}
\frac{U_{film}}{U_{roll}} \sim \frac{\sigma h_{film}}{\rho_a U_a^2 D_0^2}  = We^{-1} \overline{h}_{film}
\end{equation}
Eqn \ref{UfilmUroll} provides necessary closure to solve for $\langle D_{film} \rangle$ in eqn \ref{rolldia}. We also mention in the passing that the substitution (as discussed previously), $\rho_l U_l^2 = \rho_a U_a^2$ is used in the above equation to write it in terms of $We$. Lastly, scaling for $\overline{h}_{rim}$ and $\overline{h}_{film}$ are identical being analogous phenomena (also see main manuscript) and can be derived from the force balance between liquid inertia, $\rho_l U_l^2$ and surface tension $\sigma/h_{film\;or\;rim}$\citep{Opfer2014, Lhuissier2012}. Such a calculation readily yields a scaling dependence of the form, $We^{-1}$ which when substituted in eqn \ref{UfilmUroll} leads to the following scaling alos shown as eqn (6) in the manuscript.
\begin{equation}\label{Dfilm}
\langle \overline{D}_{film} \rangle \sim We^{-2}
\end{equation}
The average rim drop size is a consequence of conservation of volume between one segment of the wavelength, $\lambda_{rim}$ of thickness, $h_{rim}$ and the expected spherical drop size, $\langle \overline{D}_{rim} \rangle$. The important thing to note here is that we consider the rim breakup to be driven by Rayleigh-Taylor instability since the number of drops generated within a wavelength of the rim, $\lambda_{rim}$ is nearly a constant (4-5). Hence, knowing that $\overline{h}_{rim}$ scales as $We^{-1}$ we obtain the following, which is equivalent to eqn (7) in the manuscript
\begin{equation}\label{Drim}
\langle \overline{D}_{rim} \rangle \sim We^{-1}
\end{equation}
\bibliographystyle{unsrtnat}
\bibliography{SI_APL_References}